\documentclass[12pt]{emulateapj} 
\usepackage{amsmath}
\usepackage{amssymb}
\usepackage{amstext}
\usepackage{graphicx}
\usepackage{epstopdf}
\usepackage{epsfig}
\usepackage{url}
\usepackage{color}
\definecolor{myColor}{rgb}{0.9,0.9,0.9}    
\begin{document}
\renewcommand\bottomfraction{.9}
\shorttitle{Model atmospheres for massive gas giants with thick clouds}
\title{Model atmospheres for massive gas giants with thick clouds: Application to the HR~8799 planets and predictions for future detections}
\author{Nikku Madhusudhan\altaffilmark{1}, Adam Burrows\altaffilmark{1}, \& Thayne Currie\altaffilmark{2}}
\altaffiltext{1}{Department of Astrophysical Sciences, Princeton University, 
Princeton, NJ 08544; {\tt nmadhu@astro.princeton.edu, burrows@astro.princeton.edu}} 
\altaffiltext{2}{NASA Goddard Space Flight Center,Greenbelt, Maryland 20771; {\tt thayne.m.currie@nasa.gov}} 
\begin{abstract}
We have generated an extensive new suite of massive giant planet atmosphere models and
used it to obtain fits to photometric data for the planets HR~8799b, c, and d. We
consider a wide range of cloudy and cloud-free models. The cloudy models incorporate
different geometrical and optical thicknesses, modal particle sizes, and metallicities.
For each planet and set of cloud parameters, we explore grids in gravity and effective
temperature, with which we determine constraints on the planet's mass and age. 
Our new models yield statistically significant fits to the data, and conclusively confirm 
that the HR 8799 planets have much thicker clouds than those required to explain data for 
typical L and T dwarfs.  Both models with 1) physically thick forsterite clouds and a 60-$\mu m$ 
modal particle size and 2) clouds made of 1 $\mu m$-sized pure iron droplets and 1\% supersaturation 
fit the data. Current data are insufficient to accurately constrain the microscopic cloud properties, such 
as composition and particle size. The range of best-estimated masses for HR 8799b, HR 8799c, and 
HR 8799d conservatively span 2--12 M$_{J}$, 6--13 M$_{J}$, and 3--11 M$_{J}$, respectively and imply coeval ages 
between $\sim$10 and $\sim$150 Myr, consistent with previously reported stellar age. 
The best-fit temperatures and gravities are slightly lower than values obtained by Currie et al. (2011) 
using even thicker cloud models. Finally, we use these models to predict the near-to-mid IR 
colors of soon-to-be imaged planets. Our models predict that planet-mass objects follow a 
locus in some near-to-mid IR color-magnitude diagrams that is clearly separable from the 
standard L/T dwarf locus for field brown dwarfs.
\end{abstract}


\section{Introduction} 

Advances in high-contrast imaging have made it possible to image young giant 
planets at wide orbital separations around host stars. Major efforts are currently 
underway targeting progressively higher contrast levels, e.g., GPI (Macintosh et al. 2006), 
SPHERE (Beuzit et al. 2008), Project 1640 (Hinkley et al. 2008,2011), SEEDS (Tamura 2009), 
and NICI (Liu et al. 2010). Current instruments are capable of detecting contrast ratios 
in the near infrared down to $\sim 10^{-5}$ (Marois et al. 2010), leading to a number 
of planet detections in recent years. The initial discovery of a giant planet around the 
low-mass cool dwarf 2M1207 (Chauvin et al. 2004 ) was followed by more recent 
discoveries of giant planets around massive stellar hosts HR~8799 (Marois et al. 2008,2010), 
Formalhaut (Kalas et al. 2008), and $\beta$ Pic (Lagrange et al. 2009). Such detections 
provide the exciting opportunity of obtaining planetary spectra (Marois et al. 2008; 
Lafreniere et al. 2009; Hinz et al. 2010; Janson et al. 2010; Bowler et al. 2010) and 
characterizing their atmospheres, in particular their temperature structures, compositions, 
and gravities. Furthermore, the planetary radii, masses, and ages can be derived from the 
atmospheric parameters (Burrows et al. 2001; Burrows, 2005). 

Extrasolar planets detected with high-contrast imaging occupy a unique region 
in orbital phase space. Planetary spectra can also be observed for transiting 
planets. However, transit surveys for planet detection are sensitive to small 
orbital separations much less than an astronomical unit (AU). Consequently, spectra of giant 
planets at separations analogous to those in the solar system ($a \geq 5$ AU) can be 
observed only with high-contrast imaging. Since currently achievable sensitivities 
can resolve only systems that are hot, hence likely young, the planets are also likely formed in situ 
at the observed separations, and provide valuable probes for models of planet formation. 

Recent photometric observations of directly imaged planets are increasingly highlighting the need 
for a new class of atmospheric models to explain the data. Interpretations of current observations are based on traditional models that were used to interpret atmospheric spectra of brown dwarfs (Burrows et al. 2001,  2006; Baraffe et al. 2003;  Saumon \& Marley, 2008). While giant planets at large orbital distances, and, hence, with low stellar irradiation, could have been expected to have had similar spectral characteristics to low temperature brown dwarfs, recent observations are proving otherwise (Marois et al. 2008 \& 2010, Bowler et al. 2010, Currie et al. 2011). 

The HR~8799 system (Marois et al. 2008,2010) with four planetary companions at wide 
orbital separations is a case in point, epitomizing the difficulty in modeling directly imaged planets. 
In the discovery paper, Marois et al. (2008) reported six-channel near-infrared photometry of HR~8799 b, c, 
and d, and derived effective temperatures ($T_{\rm eff}$), masses ($M$), and radii ($R_p$) from 
the observed luminosities and an assumed common age of the objects, using already published 
theoretical cooling tracks (Baraffe et al. 2003). On the other hand, deriving the atmospheric parameters from spectral fits to their photometry, instead of cooling tracks, yielded unrealistically small radii (see {\it Supplementary Online Material} of Marois et al. 2008). This discrepancy was attributed to missing atmospheric physics, including inadequate cloud modeling, suggesting the need for new spectral models. 

Several follow-up observations of the HR~8799 system have also highlighted the need for new models. 
Bowler et al. (2010) reported spectroscopy of HR 8799b in the 2.12 - 2.23 $\micron$ range. While existing 
models were unable to explain all their observations, the better fits were obtained with models with thick 
clouds. Furthermore, the $T_{\rm eff}$ and $R_p$ they derived from the best fitting spectral model were 
inconsistent with those derived based on evolutionary models. Hinz et al. (2010) reported photometry 
at 3.3, 3.8, and 4.8 $\micron$, and suggested that models with non-equilibrium chemistry provided better 
fits to their data, while noting the differences in fits based on different model assumptions. 
Based on VLT spectroscopy in the 3.88-4.1 $\micron$ range (Janson et al. 2010), similar conclusions were drawn for HR 8799c. 

Currie et al. (2011) further explored the HR 8799 planets' atmospheres with data from Marois et al.(2008), 
new Subaru and VLT data, and re-reduced MMT data first presented in Hinz et al. (2010).  Using the model 
formalism of Burrows et al. (2006), they found that standard L/T dwarf atmosphere models over-predicted 
the flux between 1 and 1.25 $\micron$. They found that ``thick cloud" models, where the cloud density tracks 
that of the gas high in the atmosphere, reproduced the suppressed 1--1.25 micron flux and, thus, provided far 
improved fits to the data. However, they suggested that these models may ``overcorrect" for the weak cloud 
coverage in the standard models, as their preliminary `partly cloudy' models fit the data better. They also 
acknowledged that their grid sampling was too coarse to precisely determine best-fit surface gravities and 
effective temperatures from the data. Thus, further refinements of the ``thick cloud" formalism with a new, 
expansive, set of atmosphere models are necessary to more accurately model data for known planet-mass 
objects and predict properties of soon-to-be imaged planets. Our present work is motivated by this need. 

In this work, we embark upon an extensive modeling effort in order to present a new grid of 
atmospheric models to aid in the characterization of directly imaged planets like those of the 
HR~8799 system. We report models with different cloud types, both in physical extent and cloud composition, 
computed over a large grid in effective temperature, gravity, and metallicity. The models span a wide range 
of possible cloud cover, from clear atmospheres to those with clouds extending all the way to the top of the 
atmosphere, and encompass models that have been known to be consistent with field brown dwarfs. 
We follow the cloud parametrization developed in Burrows et al. (2006). Other studies in 
the past have also considered alternate cloud models incorporating a wide range of geometrical and optical thicknesses (e.g. Marley et al. 2002; Tsuji 2002, 2005; Helling et al. 2011).

To demonstrate the accuracy of our models, we use the HR~8799 planetary system as a test case.
We explore the model parameter space in search of the best fitting models to the photometric observations of 
HR~8799b, c, and d, while also discussing in detail the degeneracies between the various model parameters. 
As anticipated in recent literature (Marois et al. 2008, Bowler et al. 2010, Currie et al. 2011), 
the observations of the HR 8799 planets indicate a potentially new class of objects with cloud 
properties unlike those of any field brown dwarfs known. We do not consider the recently discovered fourth companion HR~8799e (Marois  et al. 2010), for which data are presently inadequate for a detailed atmospheric analysis. 
 
We report constraints on the atmospheric properties, masses, and ages of HR~8799b, c, and d. 
We use existing observations of near-infrared photometry for these objects compiled in Currie et al. (2011). 
Our best-fit models allow us to constrain $T_{\rm eff}$, $\log_{10}(g)$, mass ($M$), 
and age ($t$) for each object. The range of estimated masses for HR 8799b, HR 8799c, and HR 8799d 
conservatively span 2--12 M$_{J}$, 6--13 M$_{J}$, and 3--11 M$_{J}$, respectively. The best fits for HR~8799b, c, and d yield $T_{\rm eff}$ of $\sim$850 K, $\sim$1000 K, and $\sim$900 K, respectively, with $\log_{10}(g)$ of 4.3, 4.2, and 3.8, respectively. The corresponding estimates of the masses are $\sim$12, $\sim$11, and $\sim$6 M$_{\rm J}$, respectively, and of the ages are $\sim$150, $\sim$65, and $\sim$20 Myr, indicating that they are indeed roughly coeval, if  the adopted age of the star is $\sim$30-160 Myr.  

The observations of all the objects can be explained with models involving forsterite clouds, 
with physically thick cloud cover, and 60-$\micron$ modal grain size. Models without clouds cannot explain 
the observations for any of the objects. The observations are not explained by canonical model spectra of L 
and T brown dwarfs (Burrows et al. 2006; Saumon \& Marley 2008) at the same $T_{\rm eff}$, indicating that the present objects might form a new class with geometrically thicker clouds than those inferred for L and T dwarfs, as also suggested in Currie et al. (2011).  The recent inference of clouds in the late T Dwarf Ross~458C (Burgasser et al. 2010) suggests that this object may be related. While the requirement of clouds at altitude is stringent, the cloud composition and particle sizes are less constrained by the data, as is the gas-phase metallicity. We find that models with Fe clouds, with 1-$\micron$ size particles and 1\% supersaturation, can also fit the infrared observations. Models including non-equilibrium chemistry do not obviate the requirement for clouds. 

After establishing the accuracy of our models in constraining the HR 8799 planet properties, we present a new suite of thick cloudy atmosphere models to facilitate characterization of future detections of similar objects. We report the physical properties and absolute photometric magnitudes, in seven near/mid infrared bandpasses. Our model grid spans a wide range in planet properties, with $T_{\rm eff}$ = 600 to 1700 K, and $\log_{10}g$ = 3.5 to 5.0, corresponding to masses between 2 and 40 M$_{\rm J}$, and ages between 4 Myr and 5 Gyr, and has cloud composition and structure similar to those of our best fit models for the HR 8799 planets.  We explore the range in color-magnitude space occupied by our models and compare them with the locations of known field L/T brown dwarfs and other substellar objects. Our models predict that objects with thick clouds, like the HR 8799 planets, will occupy a distinguishable locus in color-magnitude  space, diverging from the L/T dwarf sequence at the L/T transition, and extending to systematically fainter and redder regions in color-magnitude diagrams. 

In what follows, we first describe the atmospheric models in section~\ref{sec:methods}, followed by a general exploration of the model parameter space in section~\ref{sec:results-gc}, in order to elucidate the dependence of model spectra on the various parameters. In section~\ref{sec:results-hr8799}, we describe the data used in this work and the fitting procedure, and present model fits to observations of HR~8799b, c, and d. Motivated by the model solutions for the HR~8799 planets, we provide a new collection of models in section~\ref{sec:new_models} to facilitate characterization of future detections of similar objects. Finally, in section~\ref{sec:conclusions}, we summarize our conclusions. 

\section{Methods}
\label{sec:methods}
To generate atmospheres for this paper we employ the code COOLTLUSTY (Hubeny 1988; Hubeny 
\& Lanz 1995; Burrows et al. 2006,2008). We assume hydrostatic, radiative, and chemical equilibrium 
(generally) and ignore any irradiation by the central star. At the orbital distances of these objects this is an extremely good approximation and breaks down only in the blue and ultraviolet portions of the spectrum. We use the equation of state of Saumon, Chabrier, \& Van Horn (1995), the chemical code (from which we obtain the molecular abundances as a function of temperature and pressure) of Burrows \& Sharp (1999) and Sharp \& Burrows (2007), and the opacities compiled in Sharp \& Burrows (2007) (see also Burrows 
et al. 2001). Our default spectral model employs 5000 frequency points from $\sim$0.35 to 300 microns. 
The major gas-phase species found in these theoretical atmospheres are H$_2$, He, H$_2$O, CO, CH$_4$, N$_2$, NH$_3$, FeH, CrH, Na I, and K I.  

Refractory silicates condense out at temperatures below $\sim$2300 K and above $\sim$1700 K.  The 
condensation of these refractories is automatically handled by the chemical code, which ``rains out" the 
associated species when the condensation curve indicates, leaving a gas phase depleted 
of the constituent elements in stochiometric ratios.  The resulting elemental abundances of the 
depleted atmospheres are then used to determine the chemical-equilibrium molecular abundances.
This procedure accounts for the settling of the refractories in a gravitational field and the 
depletion at altitude.  However, clouds of these refractories form and assume shapes and thicknesses that 
reflect the meteorology of the atmosphere, in particular influenced by convection.  The 
resulting cloud structures and grain properties (e.g., particle sizes) are unknown and are the most difficult 
feature of brown dwarf and hot giant exoplanet atmosphere modeling. The thickness of the 
clouds likely scales in some way with the pressure scale height, but how is almost unconstrained
by theory (Ackerman \& Marley 2001; Helling et al. 2001,2004). 

Therefore, to model clouds we parametrize the spatial extent and particle size distributions. 
The approach we use for each is taken from Burrows et al. (2006) and Sudarsky et al. (2000), 
respectively.  Because numerous refractory species condense out over a range of
temperatures and the base of a cloud is likely to intersect the associated condensation curve 
(``Clausius-Clapeyron" equation), we put the base of the clouds we study here at 2300 K (approximately 
where the most refractory species condense). At the cloud base, it is assumed that the elemental 
abundances of silicon and magnesium (dominant elements in silicate refractories) or iron are solar, or 
when non-solar atmospheres are considered that they scale accordingly. Similarly, when we focus 
on iron clouds, the elemental abundance at the base of an iron cloud is assumed to be continuous across 
the cloud/gas interface. The upward extent in pressure from this base is then parametrized 
in terms of some multiple or fraction of the local pressure scale height. The baseline distribution 
of the cloud particle density is assumed to follow the gas-phase pressure profile and we multiply 
this baseline reference distribution by a shape function, $f(P)$, to determine the actual model cloud 
distribution in pressure space (hence, real space). $f(P)$ is never greater than one and can be made 
to cut off sharply at the top and bottom of the cloud. Specifically, we define $f(P)$ as:
\begin{equation}
f(P) = 
\begin{cases}
\left(P/P_{\rm u}\right)^{s_{\rm u}}, & P \le P_{\rm u} \\
f_{\rm cloud}, & P_{\rm u} \le P \le P_{\rm d} \\
\left(P/P_{\rm d}\right)^{-s_{\rm d}}, & P \ge P_{\rm d} \, , \\
\end{cases}
\label{eq:cloud}
\end{equation}
where $P_{\rm d}$ is the pressure at the cloud base and $P_{\rm u} < P_{\rm d}$.  In particular, 
$P_{\rm u}$ is the pressure at the intersection of the atmospheric $T/P$ profile and the 
condensation curve of the least refractory condensates, assumed here to be either forsterite or iron. 
The indices ${s_{\rm u}}$ and ${s_{\rm d}}$ define the rapidity with which the clouds are cut off
on their upper and lower boundaries.  The larger they are, the sharper the cutoff.
When $f_{\rm cloud} = 1$ the cloud has a flat portion in its middle between $P_{\rm u}$ and $P_{\rm d}$.
The parameters ${s_{\rm u}}$, ${s_{\rm d}}$, and $f_{\rm cloud}$ define the cloud spatial structure.

For this study, we consider four model parameter sets:
\begin{equation}
\begin{split}
\mbox{Model E: } s_{\rm u}=6, s_{\rm d}=10, f_{\rm cloud}=1\\
\mbox{Model A: } s_{\rm u}=0, s_{\rm d}=10, f_{\rm cloud}=1\\
\mbox{Model AE: } s_{\rm u}=1, s_{\rm d}=10, f_{\rm cloud}=1\\
\mbox{Model AEE: } s_{\rm u}=2, s_{\rm d}=10, f_{\rm cloud}=1.\\
\end{split}
\end{equation}
Since $s_{\rm d}=10$, all these models have a sharp cutoff at high pressures (at their base).
Also, all these cloud models have a ``flat" profile in the middle that follows the pressure profile itself.
The latter is roughly exponential with the pressure scale height.  It is in the extent of the cloud at altitude
that these models differ.  Model E cuts off rapidly at altitude.  Model A extends all the way to the top of the atmosphere,
with a particle number density distribution that follows the gas.  Model AE is intermediate between Models E and A,
decaying at the cloud tops with a scale height that is equal to twice that of the gas.  Nevertheless, it still extends 
beyond Model E and puts particles higher up at lower pressures. Model AEE is similar to Model AE, but extends 
to altitude with a scale height of a third of a pressure scale height.  Model E was the model settled upon in 
Burrows et al. (2006) as the best fitting model for L and T dwarfs, though it still did not capture the rapidity 
of the L-to-T transition. To model the HR8799 planets, we are forced to extend the cloud tops to lower pressures, 
thickening the clouds to a specified degree. We have thus introduced the `AE' cloud models in this work. 

We refer to the AE-type clouds as `thick clouds' in this work, implying that they are thicker 
in physical extent than the E-type clouds that are known to fit observations of L dwarfs. For the same grain 
size distribution, the physical thickness naturally also enhances the optical thickness of the cloud. Our goal in the 
present work is to constrain the combination of physical extent and cloud grain properties of clouds that best explain 
the observations of the HR~8799 planets. Recent studies have advocated the need for such thick cloudy models 
to explain the HR~8799 observations (Bowler et al. 2010, Currie et al. 2011). Several theoretical studies in the past 
have also considered cloud properties comparable to our thick cloud models (e.g. Chabrier et al. 2000; Allard et al. 2001; Tsuji 2002 \& 2005; Burrows et al. 2006). 

For the size distribution of cloud particles, we use the Deirmendjian distribution of cumulus water clouds on Earth 
(Sudarsky et al. 2000):
\begin{equation}
\frac{dN}{da} \propto \left( \frac{a}{a_0} \right)^6 \exp{\left[ -6 \left(\frac{a}{a_0} \right) \right]}.
\end{equation}
The modal particle size is $a_0$ and is a free parameter in our model.  
In this paper, we assume spherical particles of either forsterite (Mg$_2$SiO$_4$) or iron, 
and use Mie scattering theory with a table of complex indices of refraction 
as a function of wavelength for each (Sudarsky et al. 2000). In Burrows et al. (2006), the fiducial best-fitting 
model had not only a Model E structure, but forsterite particles with a modal size of 100 microns.  Since iron 
undergoes homogeneous condensation (i.e. vapor and liquid/solid phases composed of the same chemical species), we assume a supersaturation factor of 1\% to represent the fraction of iron 
in the condensed phase (Fegley \& Lodders, 1996; Ackerman \& Marley 2001; Marley et al. 2002; 
Burrows et al. 2006). As mentioned above, the iron elemental abundance at the base of an iron cloud is 
assumed to be continuous across the cloud/gas interface. Any elemental discontinuities would be smoothed 
out by upward gaseous diffusion and overshoot from the lower, hotter, regions of the atmosphere.

The opacity spectrum of forsterite clouds is similar to that of enstatite clouds in mid-IR wavelengths 
($\lambda \gtrsim 5 \micron$). In the near-IR, the forsterite clouds have higher opacity in the 1$-$3 $\micron$ 
range compared to enstatite clouds. On the other hand, the opacity spectrum of Fe clouds is significantly 
different from both enstatite and forsterite clouds. The Fe clouds offer substantially higher grain opacities 
in the near-infrared, with a steep decline into the mid-infrared. We do not consider in detail 
the dependence of cloud formation processes on the gas density and gravity, although cloud formation is 
favored at higher densities, and the microscopic nature of clouds in the different gravity regimes could be 
different.

\section{Dependence of Predicted Spectra on Model Parameters}
\label{sec:results-gc}

Before describing detailed model fits to observations of the HR~8799 planets, we present a general exploration of the parameter 
space to understand some generic features of model spectra and their dependence on model parameters. 
The parameters in our model are: $T_{\rm eff}$, $\log_{10}(g)$, metallicity, cloud composition, modal particle 
size, and the physical extent of cloud cover. We also explore models of non-equilibrium chemistry, in which case 
the eddy diffusion coefficient, $K_{zz}$, (Saumon et al. 2003; Hubeny \& Burrows, 2007) is an additional parameter. 

\subsection{Cloud-free models}
\label{sec-cloudfree}
\begin{figure}[ht]
\centering
\includegraphics[width = \textwidth]{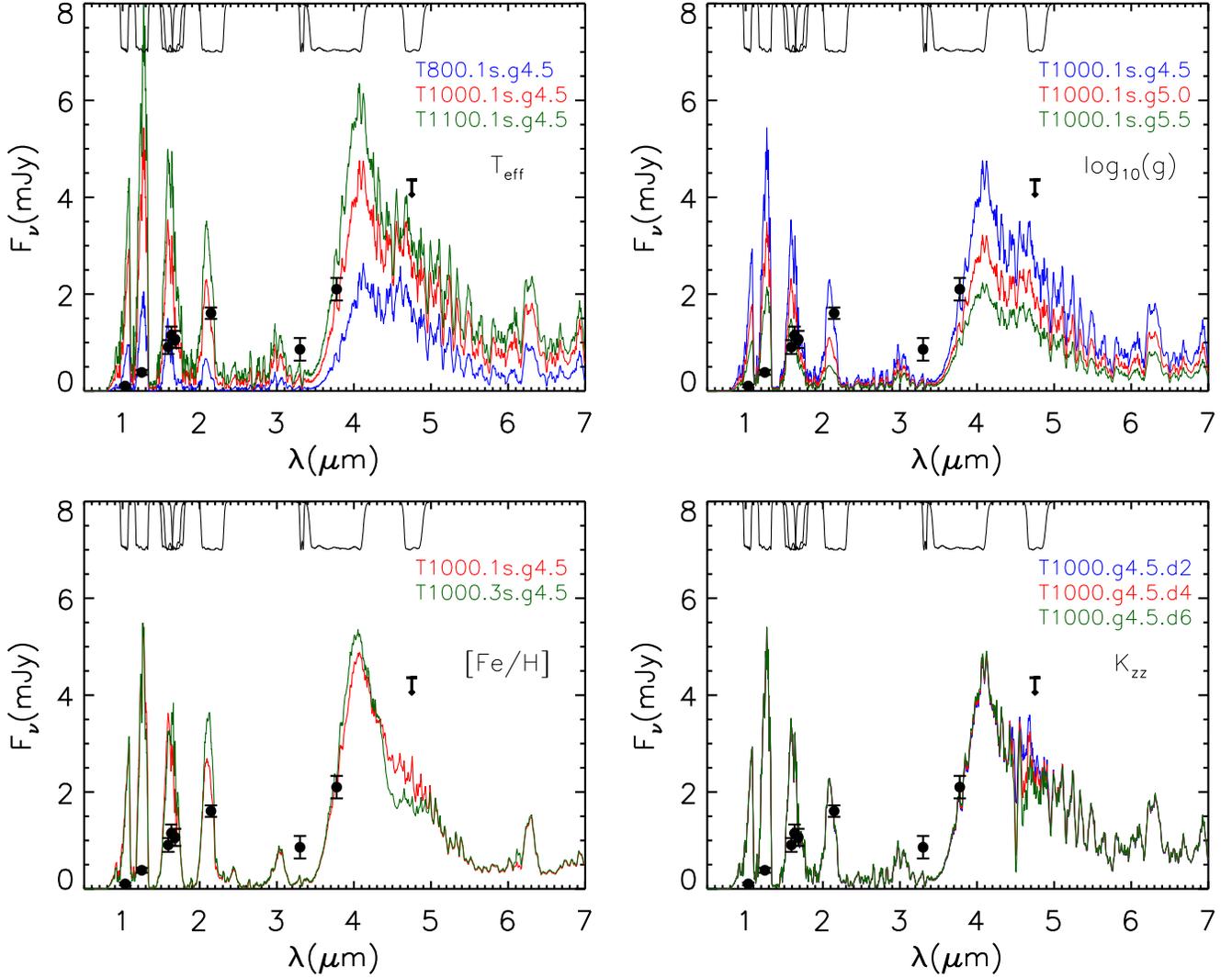}
\caption{Model spectra of giant planet atmospheres without clouds. The solid lines show model spectra, 
and photometric observations of HR 8799b are shown in black. The observations are described in 
section~\ref{sec:observations}. Each panel shows a range of models exploring a particular parameter: 
$T_{\rm eff}$, $\log_{10}(g)$, [Fe/H], and the eddy diffusion coefficient $K_{\rm zz}$. The models in the 
top-left, top-right, and bottom-left panels assume equilibrium chemistry with the specified abundances.
The models in the bottom-right panel have non-equilibrium CO-CH$_4$ chemistry, as described in 
section~\ref{sec-cloudfree} and in Hubeny \& Burrows (2007). The model names in the legend specify the parameters values. 
For example, in the top-left panel, T800.1s.g4.5 denotes a  cloud-free (`T') model with $T_{\rm eff}$ = 800 K, $\log_{10}(g) = 4.5$, 
and 1$\times$solar metallicity. For the models in the bottom-right panel, the abundances are solar, 
and the legend specifies the diffusion coefficient; for example, `d6' implies $K_{\rm zz} = 10^6$ cm$^2$ 
s$^{-1}$. 
Cloud-free models predict $\sim$2-5 times higher fluxes in the z, J, and H 
bands than are observed for HR 8799b. All three objects considered in this study, HR~8799b, c, and 
d, require thick clouds to explain the observations.} 
\label{fig:clear}
\end{figure}

We first consider model atmospheres without clouds. Figure~\ref{fig:clear} shows cloud-free models as a 
function of the various model parameters, along with observations of HR~8799b, shown for reference. 
Here, we focus on a nominal spectrum with $T_{\rm eff} = 1000$ K and $\log_{10}g = 4.5$, the red model 
in  Fig.~\ref{fig:clear}. The model nomenclature specifies the atmospheric parameters. For example, `T1000.1s.g4.5' refers to a cloud-free (`T') model with $T_{\rm eff}$ = 1000 K, $\log_{10}(g) = 4.5$, and 1$\times$solar metallicity. The major features in the spectrum can be understood in terms of 
the dominant opacity sources. The high fluxes in the J, H, and K bands (within $\sim$1 $-$ 2.4 $\micron$) are due to windows in water vapor opacity, and are a defining characteristic of cloud-free models; the gaps 
between the high fluxes correspond to absorption bands of water vapor. The deep absorption trough 
between  2.2 and 3.5 $\micron$ is due to water and methane absorption. Water absorption also causes 
deep features between 5 and 7 $\micron$, while the peak between 4 and 5 $\micron$ is due to the lack 
of significant absorption features due to any of the dominant species. In principle, CO has strong absorption 
features between 4 and 5 $\micron$. However, for $T_{\rm eff} = 1000$ K, and assuming equilibrium chemistry, methane is the dominant carbon carrier, as opposed to CO. 

The effects of $T_{\rm eff}$, $g$, metallicity, and non-equilibrium chemistry on model spectra of cloudless 
atmospheres are shown in Fig.~\ref{fig:clear}. An increase in $T_{\rm eff}$ causes an increase in emergent 
flux across all wavelengths. However, the relative increase in flux varies with wavelength, depending on the molecular absorption. In regions of low water opacity, such as in J, H, and K, and between 4 and 5 $\micron$, the emergent flux is highly sensitive to temperature. On the other hand, the emergent flux is less 
sensitive to temperature at wavelengths where the spectrum is saturated with water absorption,  as shown 
in the top-left panel of Fig.~\ref{fig:clear}. It is also apparent that none of the cloudless models can explain the 
observations of HR 8799b, shown for reference, at wavelengths below 2.2 $\micron$. 

The dependence of spectra on gravity is shown in the top-right panel of Fig.~\ref{fig:clear}.  A decrease in 
gravity causes a general increase in flux across all wavelengths, except in regions saturated by strong 
water vapor absorption. The dependence of the spectra on gravity arises from two factors. For a given 
$T_{\rm eff}$, a lower gravity causes an increase in radius (see Eq. 5 of Burrows et al. 2001), leading 
to a corresponding increase in the emergent flux. In addition, a higher gravity causes a higher mean density in the atmosphere, leading to higher opacities, especially in the H$_2$-H$_2$ collision-induced  opacity, 
which increases with density. The higher opacities then cause enhanced flux suppression in 
all but the 3.3-$\micron$ channel which is saturated with methane and water opacity. 

At the level of present observations, and in the context of our models, the spectra are not particularly sensitive to changes in metallicity or to non-equilibrium chemistry due to eddy mixing. As shown in the bottom-left panel of 
Fig.~\ref{fig:clear}, variation in metallicity between solar and 3$\times$solar values, does not significantly alter the spectra in the observed bands. Finally, as shown in the bottom-right panel of Fig.~\ref{fig:clear}, non-equilibrium chemistry induces nominal changes in the model spectra, albeit too weak to be determined with current observations. For models with non-equilibrium chemistry, the eddy diffusion coefficient ($K_{zz}$) causes transport of CO from the hotter lower atmosphere to the upper, observable regions of the atmosphere (Saumon et al. 2003; Hubeny \& Burrows  2007). Consequently, higher $K_{zz}$ and faster diffusion leads to more CO absorption, which is apparent in the pronounced absorption features in the 4 - 5 $\micron$ spectral region. 

\subsection{Cloudy models}
\label{sec-cloudy}

\begin{figure}[ht]
\centering
\includegraphics[width = \textwidth]{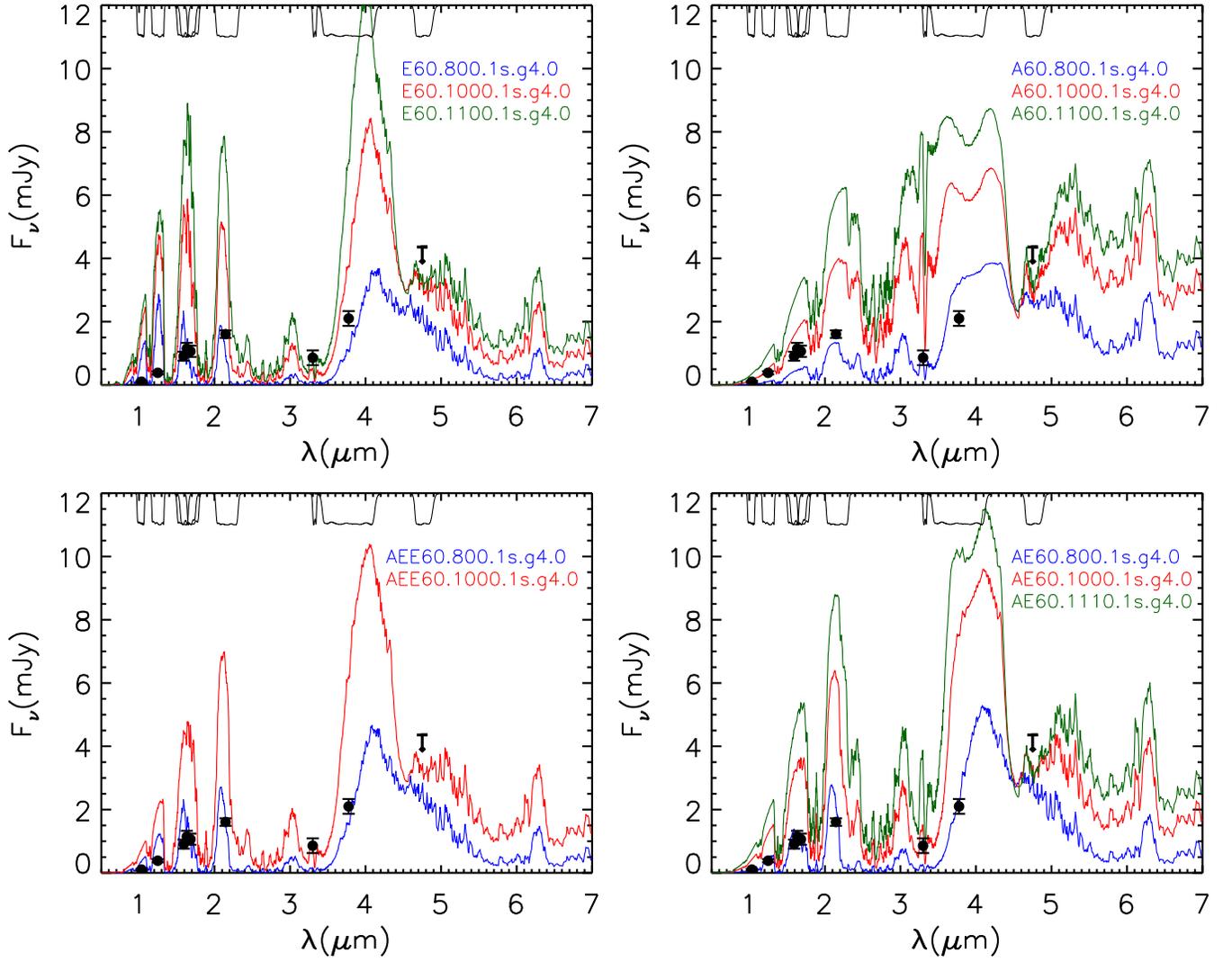}
\caption{Model spectra of giant planet atmospheres with clouds.  Four types of cloudy models, as described 
in section~\ref{sec:methods}, are shown. The top-left panel shows models with E-type clouds which truncate at high altitudes. The top-right panel shows models with A-type 
clouds which extend up to the top of the atmosphere, and have the thickest extent among the models 
considered in this work. The bottom panels show AEE and AE cloud types, which are intermediate between 
the A and E types, as described in section~\ref{sec:methods}. The nomenclature E60.800.1s.g4.0 stands for 
E-type clouds with 60-$\micron$ modal particle size, $T_{\rm eff}$ = 800 K, 1$\times$solar metallicity, and 
$\log_{10}g$ = 4.0. Observations of HR~8799b are shown in black. AE-type clouds provide the 
best match to the observations (see Fig.~\ref{fig:spectra_b} and discussion in section~\ref{sec-hr8799b}).}
\label{fig:clouds}
\end{figure}

Model spectra of atmospheres with clouds are shown in Fig.~\ref{fig:clouds}. Four types of cloudy 
models (E, A, AEE, and AE) are shown, along with observations of HR~8799b, shown for reference. As described 
in section~\ref{sec:methods}, the A models have cloud cover extending all the way to the top of the atmosphere, 
whereas the E models cut off sharply at altitude. The AE and AEE models are intermediate between A and E models, with the AEE models leaning more towards the E models for the same $T_{\rm eff}$ and $g$. All 
the models have equilibrium chemistry with solar abundances. The model nomenclature encapsulates the 
cloud type and atmospheric parameters. For example, `E60.1000.1s.g4.0' implies `E' type cloud cover with 
60-$\micron$ modal particle size, $T_{\rm eff} = 1000$ K, 1$\times$ solar metallicity, and $\log_{10}g = 4.0$, 
where $g$ is in cm s$^{-2}$. 

Model spectra of atmospheres with `E' type clouds are shown in the top-left panel of Fig.~\ref{fig:clouds}. 
E-type models have been shown to provide good fits to observations 
of L and T dwarfs (Burrows et al. 2006). Because E-type models have minimal cloud cover among the models 
considered here, their spectral features bear some resemblance to those of cloud-free models shown in 
Fig.~\ref{fig:clear}. The large peaks in the z, J, H, and K bands show regions of low water opacity, 
as also seen in the cloud-free models. But, the peaks can be substantially lower, and troughs higher, 
compared with those of cloud-free models, at high $T_{\rm eff}$ and $g$ (Burrows et al. 2006). In the present 
context, however, the model fluxes in the z, J, and H bands are still too high by a factor of 2 - 5 
than the observations of the HR~8799 objects in these bandpasses, as suggested by data for HR~8799b 
in Fig.~\ref{fig:clouds}. 

Clouds of thicker physical extent than the E models cause greater suppression of the emergent 
flux in the z, J, and H bands. This situation is seen in models with A-type clouds, as shown 
in the top-right panel of Fig.~\ref{fig:clouds}. While A models can, in principle, yield low observed fluxes 
in the z, J, and H bands, they predict much higher fluxes than E models at longer 
wavelengths, in the L and M bands, for the same  $T_{\rm eff}$. This is caused by the spectral 
redistribution of energy from the shorter wavelengths, where cloud opacity suppresses the 
emergent flux, to longer wavelengths in the spectrum. However, as shown in the top-right panel of 
Fig.~\ref{fig:clouds}, A models cannot fit all the observations of HR~8799b simultaneously. 

\begin{figure}[ht]
\centering
\includegraphics[width = \textwidth]{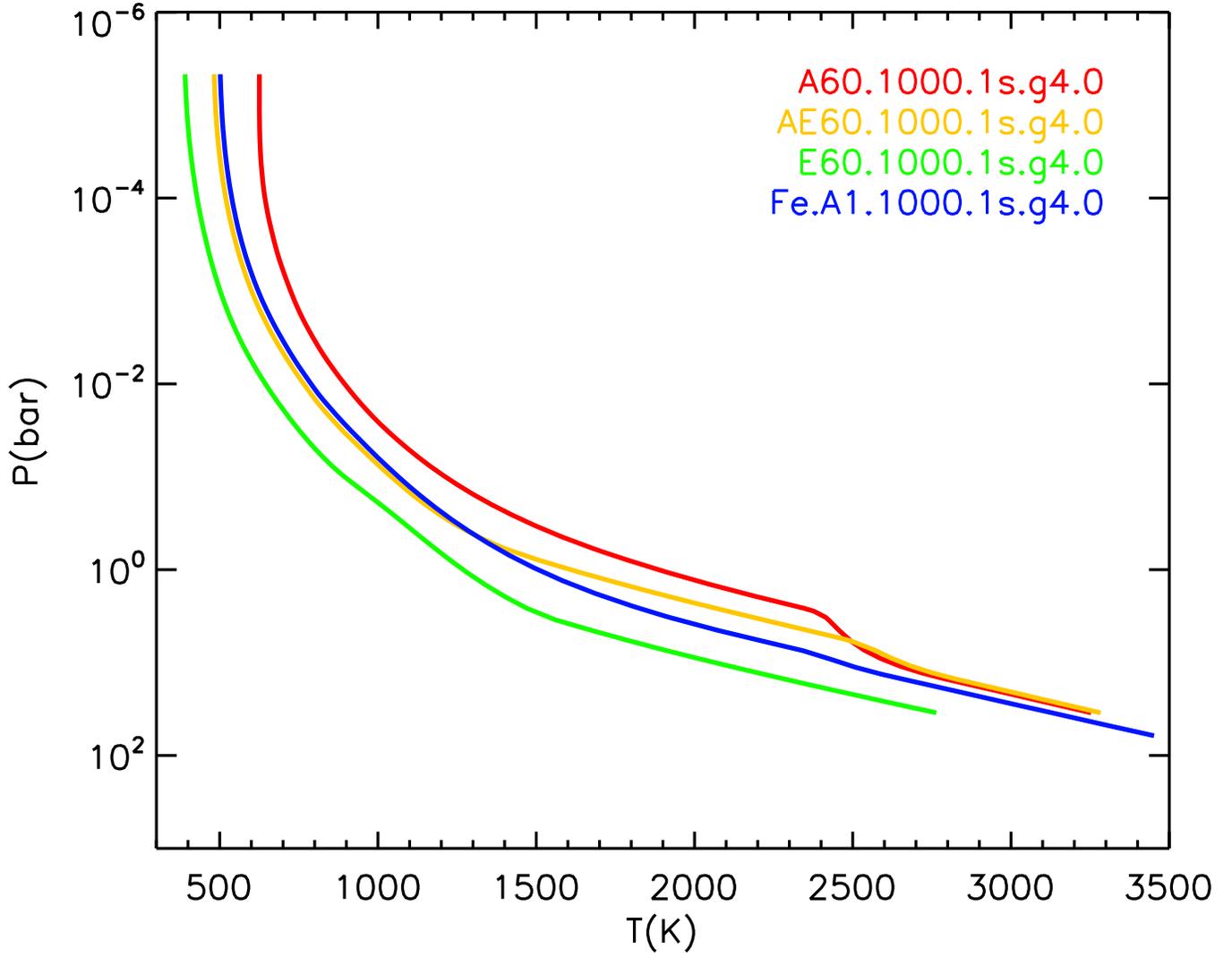}
\caption{Temperature-Pressure ($T$-$P$) profiles of model atmospheres with clouds. The model nomenclature 
shown in the legend is explained in Fig.~\ref{fig:clouds} and in section~\ref{sec:methods}. All the models 
have $T_{\rm eff}$ = 1000 K, $\log_{10}g$ = 4.0, and solar metallicity, but different cloud types. The A-type 
(E-type) forsterite clouds which have the thickest (thinnest) cloud cover have the hottest (coolest) $T$-$P$ 
profiles for the same $T_{\rm eff}$. Models with AE-type forsterite clouds with 60-$\micron$ modal particle 
size and those with A-type iron clouds with 1-$\micron$ modal particle size, with 1\% supersaturation, have similar 
$T$-$P$ profiles.}
\label{fig:tp}
\end{figure}

The above discussion suggests that  cloudy models with physical extent intermediate between A and E 
models might be able to explain the data better. The bottom-left and bottom-right panels of Fig.~\ref{fig:clouds} 
show such models, i.e., the AEE and AE models described in section~\ref{sec:methods}. While AEE models, 
those that are closer to the E models, are still inadequate to explain the observations, the AE models provide 
the best match to the data. The thicker physical extent of the clouds in the A and AE models over 
those of the E model are apparent in the total optical depths contributed by each to the observable atmosphere. 
For example, considering the models in Fig.~\ref{fig:clouds} with $T_{\rm eff} = 1000$K and $\log_{10}g = 4.0$, the total Rosseland optical depths at pressures below 1 bar for the A, AE and E models are 12.6, 5.0, and 1.2, respectively. Considering pressures below 0.3 bar, the total Rosseland optical depths for the same A, AE, and E models are 4.4, 1.1, and 0.6, respectively. In what follows, we present the best fits to the observations of the three objects HR~8799b, c, and d, based on AE models. Figure~\ref{fig:tp} shows temperature-pressure ($T$--$P$) profiles for four representative cloudy models. A-type forsterite cloud models which have the thickest (physically) cloud cover require hotter temperature profiles, for the same $T_{\rm eff} = 1000$ K and $\log_{10}g = 4.0$, compared to AE-type and E-type models. On the other hand, AE-type forsterite clouds and A-type iron clouds, with 1\% supersaturation, yield similar $T$--$P$ profiles. 

\section{Application to the HR 8799 planets}
\label{sec:results-hr8799}
In this section, we present model spectra that fit the observations of HR~8799b, c, and d, along 
with estimates of masses and ages. For each object, we have extensively explored the parameter 
space of models with and without clouds in search of the best-fitting models. As described in 
section~\ref{sec:results-gc} for the case of HR~8799b, we find that cloud-free models cannot 
explain the data for any of the three objects, and that cloudy models with AE-type clouds 
provide the best match to the data. Below, we present the constraints on the $T_{\rm eff}$, 
$\log_{10}g$, mass (M), and age (t) for each object, along with the best fitting models. 

\subsection{Observations and Fitting procedure} 
\label{sec:observations}
Several recent studies of the HR~8799 system have reported broadband photometry and 
colors of the companions. For the spectral fits in the present study, we use broadband photometry 
at nine wavelengths reported in the literature for HR~8799 b, c, and d. The observations we use are 
described and summarized in Currie et al. (2011). The nine filters and their central wavelengths 
are: z (1.03 $\micron$)\footnote{The Subaru/IRCS z band filter used by Currie et al. (2011) is very similar to 
the better known MKO ``Y" band filter. The IRCS z filter and Y filter should not be confused with the Sloan 
z$^\prime$ filter, which covers shorter wavelengths.}, J (1.248 $\micron$), H (1.633 $\micron$), CH4$_S$ (1.592 $\micron$), CH4$_L$ (1.681 $\micron$), K$_s$ (2.146 $\micron$), [3.3] (3.3 $\micron$), L$^\prime$ (3.776 $\micron$), 
and M (4.75 $\micron$). The photometry in the J, H, CH4$_S$, CH4$_L$, and K$_s$ bands for all 
the three objects is taken from Marois et al. (2008). The observations in the [3.3], L$^\prime$, and M 
bands are adopted from Currie et al.~(2011). The observations in the M-band for all the three objects 
are 3-$\sigma$ upper limits, as are the z-band observations of HR 8799c and d, and the [3.3]-band observation of HR 8799d. We do not consider the recently discovered fourth companion HR~8799e (Marois et al. 2010; 
Currie et al. 2011) as data for this planet are currently restricted to Ks and L$^\prime$ band and thus are 
inadequate to place meaningful constraints on its atmospheric parameters. 

We determine the best-fit models to observations of each object using a grid-based approach. 
We computed models over a large grid in $T_{\rm eff}$ and $\log_{10}g$ for different cloud types, particle 
sizes, and metallicities. After a global exploration of the grid, we found that AE type cloudy models provided 
the best match to the data. Consequently, we report error contours in a two-dimensional 
($T_{\rm eff}$ -- $\log_{10}g$) grid of AE-type forsterite cloud models. We nominally fix the modal particle 
size at 60 $\micron$ and the metallicity at solar, but we found that forsterite particle sizes of 
30 -- 100 $\micron$ provided almost equally good fits. We evaluate the model fits using the 
$\chi^2$ statistic, given by:
\begin{equation}
\chi^2 = \sum_{i = 1}^{N_{obs}} \bigg(\frac{f_{i,model} - f_{i,obs}}
{\sigma_{i,obs}} \bigg)^2,
\label{eq:xi}
\end{equation}
where, $f_{i,model}$ is the channel-integrated emergent flux of the model in each channel, 
and $f_{i,obs}$ and $\sigma_{i,obs}$ are the observed flux and the 1-$\sigma$ 
uncertainty in that channel, respectively. $N_{obs}$ is the number of observations. 
Where only 3$\sigma$ upper limits are available, we compute their contribution to the $\chi^2$ using 
the following approach. When the model flux is less than the upper limit, we 
do not penalize the $\chi^2$. When the model flux exceeds the upper limit, the contribution to 
the $\chi^2$ is $(3[f_{model} - f_+]/f_+ )^2$, where $f_+$ is the observed 3$\sigma$ upper limit.  

We derive the confidence limits on the model fits based on the $\chi^2$ distribution and the 
number of degrees of freedom ($N_{dof}$). When $N_{obs} = 9$ observations $(N_{dof} = 8)$ 
are used in computing the $\chi^2$, the $3\sigma$, $4\sigma$, $5\sigma$, $6\sigma$, and 
$7\sigma$ confidence levels correspond to $\chi^2$ of 23.6, 33.0, 44.0, 56.8, and 71.1, respectively.  
For $N_{obs} = 8$, i.e. $N_{dof} = 7$, the corresponding values of $\chi^2$ are 
21.8, 31.0, 41.8, 54.4, and 68.5, respectively. 

\subsection{HR 8799b}
\label{sec-hr8799b}

\begin{figure}[ht]
\centering
\includegraphics[width = 0.75\textwidth]{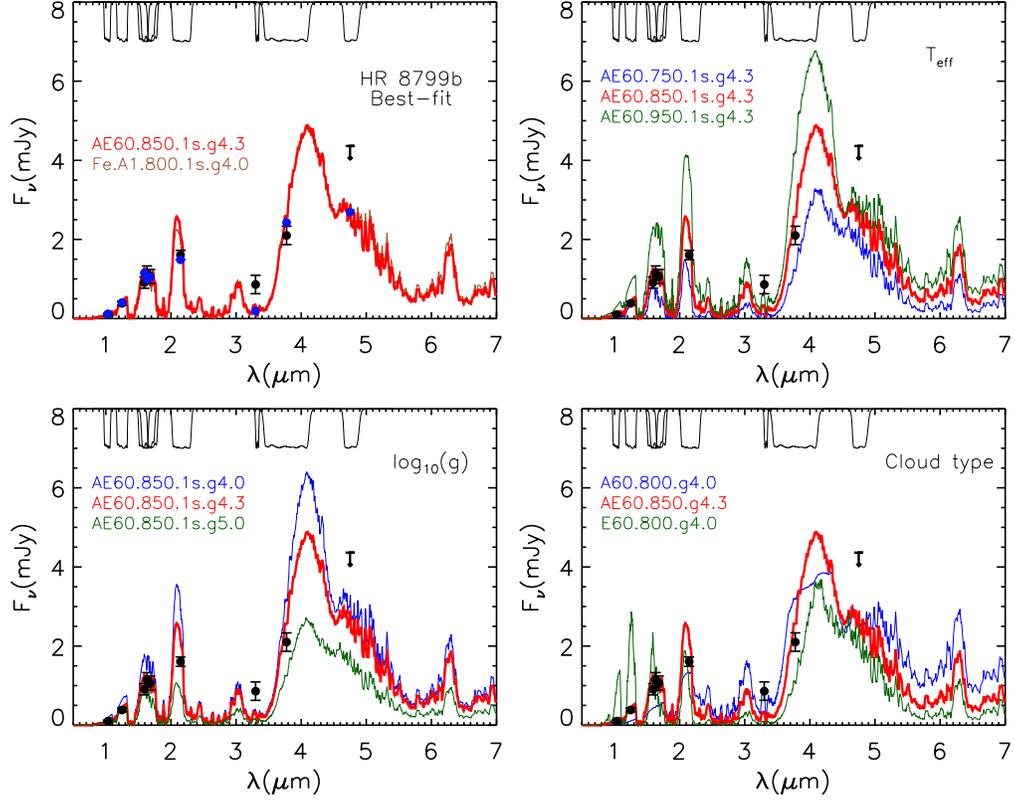}
\caption{Model spectra of HR~8799b. The observations, shown in black, are from the literature (described in 
section~\ref{sec:observations}). The model nomenclature is described in Fig.~\ref{fig:clouds} and in section~\ref{sec-hr8799b}. The black lines 
at the top of each panel indicate the photometric bandpasses, and the colored solid lines show the model 
spectra. The top-left panel shows the best-fit models. The observations are best explained by cloudy models 
with intermediate cloud cover (AE type), with modal particle size of 60 $\micron$, 
$T_{\rm eff}$ = 850 K, metallicity of 1$\times$ solar, and $\log_{10}(g) = 4.3$, shown in the red curve.  The blue 
filled circles show the best-fit model integrated over the photometric bandpasses. The brown curve in the top-left panel shows a model with Fe clouds with 1\% supersaturation, and 1-$\micron$ particle size, with $T_{\rm eff}$ = 800 K, metallicity of 1$\times$ solar, and $\log_{10}(g) = 4.0$, which also provides a good fit to the observations, showing that different cloud types and compositions can in principle explain the photometric points. The remaining panels show 
departures from the best-fit model (red model in top-left panel) in temperature (top-right), gravity (bottom-left), 
and cloud type (bottom-right). The values of the varied parameter for the different models are  shown in the 
legend; all unspecified parameters are the same as for the best-fit model. Higher $T_{\rm eff}$ yields systematically 
higher fluxes in all the channels, except at 3.3 $\micron$. E-type cloud models which explain the spectra of L and 
T dwarfs fairly well are inadequate  to explain the current observations. For a given $T_{\rm eff}$, E-type cloud 
models predict 2--5 times higher fluxes at shorter wavelengths (in z, J, and H bands), than those required by the data. On the other hand, purely A models yield excess flux 
at short wavelengths and low flux at longer wavelengths. The intermediate AE-type models explain the data best. 
Changing the metallicity does not improve the fits significantly. See text (section~\ref{sec-hr8799b}) for discussion.}
\label{fig:spectra_b}
\end{figure}

Based on a detailed exploration of cloudy models, we find that the observations of HR~8799b are best 
explained by models with forsterite clouds with physically thick AE-type cloud cover, and a nominal 
modal particle size of 60 $\micron$. As described in section~\ref{sec:methods}, AE-type cloud models have cloud 
cover that is intermediate between the A and E type cloud models. As shown in the top-left panel of 
Fig.~\ref{fig:spectra_b}, the observations can be explained with an AE-type forsterite cloud model 
with $T_{\rm eff}$ = 850 K and $\log_{10}g$ = 4.3. The observations can also be explained by models 
with thick A-type Fe clouds with 1-$\micron$ modal particle size and 1\% supersaturation, implying multiple possibilities for the cloud composition and particle size. In the 
remaining three panels of Fig.~\ref{fig:spectra_b}, we show the dependence of our best-fit model 
on $T_{\rm eff}$, $\log_{10}g$, and cloud properties. An increase in  $T_{\rm eff}$ or a decrease in 
$g$ over the best-fit model leads to systematically higher fluxes across the spectrum, as described in 
section~\ref{sec:results-gc}. The bottom-right panel shows the dependence on cloud-cover type. 
A-type clouds, with complete cloud cover up to the top of the atmosphere, cause excess 
suppression of flux in the 1-2 $\micron$ range. On the other hand, E-type clouds that have 
cloud cover truncated at height are insufficient to suppress the 1-2 $\micron$ flux to the 
observed levels. The degeneracy in cloud composition, between forsterite and Fe clouds, might 
be resolved by future observations in the visible and mid-infrared. Iron clouds have strong 
absorption in the visible, whereas silicates have strong spectral features at $\sim$10 $\micron$ 
and $\sim$20 $\micron$ which might be observable in high resolution spectra (Cushing et al. 2006; 
Helling et al. 2006; Henning 2010).

\begin{figure}[ht]
\centering
\includegraphics[width = 0.75\textwidth]{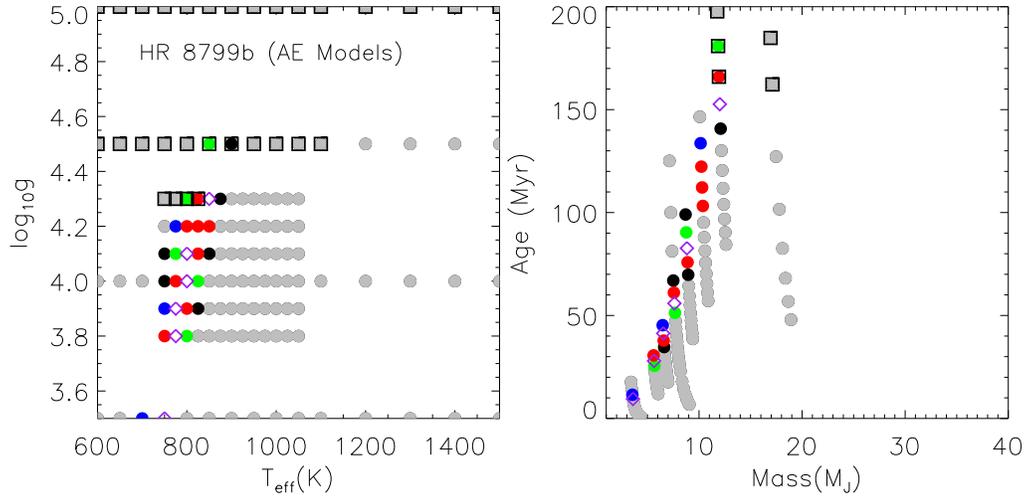}
\caption{Model space explored for HR~8799b. The symbols show models with AE clouds 
computed over a grid in $T_{\rm eff}$  and $\log_{10}(g)$, as shown in the left panel. The right panel 
shows the corresponding regions in the Mass-Age plane, derived from $T_{\rm eff}$  and $\log_{10}(g)$ (Burrows et al. 
2001). The models are color coded by the degree of fit to the data, quantified by the confidence level associated with 
the $\chi^2$ (see section~\ref{sec:observations}). The $\chi^2$ is derived from fits to photometric data in all the nine 
wavebands available. The purple open-diamonds show fits to within the $3\sigma$ confidence level, and the red, blue, 
green, black, and gray filled-circles indicate fits within $4\sigma$, $5\sigma$, $6\sigma$, $7\sigma$, and $>7\sigma$, 
confidence levels, respectively. The colored points enclosed in black boxes indicate models that yield an age for the 
object that is greater than the reported maximum age of the star (160 Myr), and, hence, are ruled out as implausible solutions despite their level of fit to the data. 
} 
\label{fig:cplot_all_b1}
\end{figure}

\begin{figure}[ht]
\centering
\includegraphics[width = 0.75\textwidth]{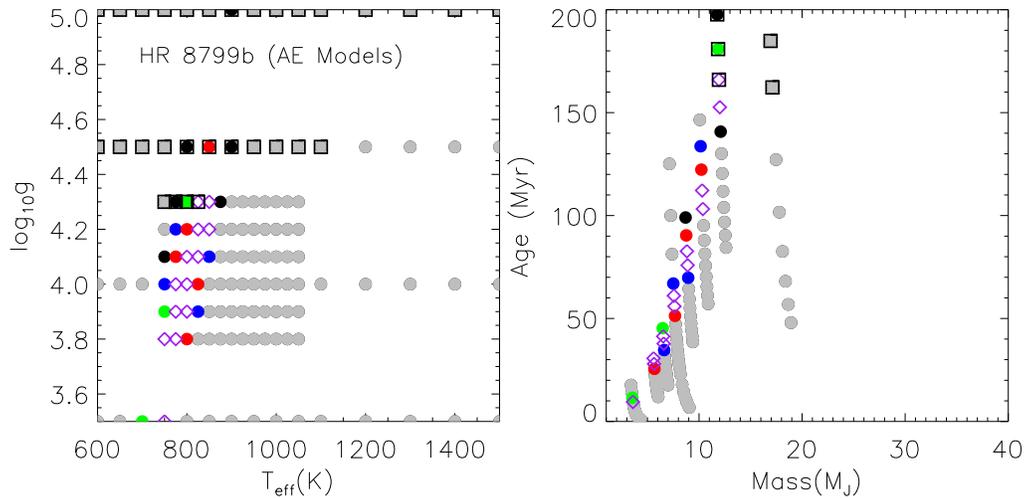}
\caption{Optimal constraints on the atmospheric parameters of HR~8799b. The description of this figure is same as that for 
Fig.~\ref{fig:cplot_all_b1}, with the exception that the 3.3-$\micron$ observation is excluded from the $\chi^2$, 
implying $N_{obs} = 8$ observations and $N_{dof} = 7$ degrees of freedom (see section~\ref{sec-hr8799b} for a 
discussion).} 
\label{fig:cplot_all_b2}
\end{figure}

The space of models explored in order to fit the observations of HR~8799b, along with distribution of 
the $\chi^2$ in this space, are shown in Figs.~\ref{fig:cplot_all_b1} and \ref{fig:cplot_all_b2}. Each model 
in the large grid explored is shown color-coded by the degree of fit to the data, denoted by the confidence 
level corresponding to the $\chi^2$. The purple open-diamonds show fits to within the $3\sigma$ confidence 
level, and the red, blue, green, black, and grey, filled-circles indicate fits within 
$4\sigma$, $5\sigma$, $6\sigma$, $7\sigma$, and $> 7\sigma$, confidence levels, respectively. 
In Fig.~\ref{fig:cplot_all_b1}, the $\chi^2$ is computed by taking into account all the nine 
photometric observations described in section~\ref{sec:observations}. In this case, we obtain a high 
$\chi^2$ for all the models, with only a few models lying within the $3\sigma$ level. We find 
that a large part of the $\chi^2$ is contributed solely by the 3.3-$\micron$ data point, as is also evident from 
Fig.~\ref{fig:spectra_b}. This discrepancy can in principle be explained by a lower methane abundance than the equilibrium abundances (at solar metallicity) considered in our cloudy models. Without the contribution of the 3.3-$\micron$ point to the $\chi^2$ we obtain much better fits to all the remaining data, while still preserving the best-fit model parameters, as shown in Fig.~\ref{fig:cplot_all_b2}. Instead of generating an entirely new 
grid of models with less methane in order to fit the single 3.3-$\micron$ point, we assess the model 
fits based on the remaining eight observations. We consider the latter procedure as a more accurate means 
of assessing the model fits to the data and quote the ranges of fit parameters accordingly. Consequently, 
we have $N_{obs} = 8$ observations, yielding $N_{dof} = 7$ degrees of freedom. 

The ranges of the physical parameters of HR~8799b at different levels of confidence are shown in 
Fig.~\ref{fig:cplot_all_b2}. The constraints at 3$\sigma$ confidence and the parameters for the best-fit 
model are shown in Table~\ref{tab:spectra_1} and Table~\ref{tab:spectra_2}. The solutions 
at 3$\sigma$ confidence yield $T_{\rm eff}$ = 750  - 850 K, $\log_{10}g$ = 3.5 - 4.3, M = 2 - 12 M$_{\rm J}$, and t = 10 - 150 Myr. The best-fitting model, reported in Table~\ref{tab:spectra_1}, has $T_{\rm eff}$ = 850 K and $\log_{10}g$ = 4.3.  From the $T_{\rm eff}$ and $\log_{10}g$, we derive a mass of $\sim$12 M$_{\rm J}$ and an age of $\sim$150 Myr, using analytic fits to evolutionary models in the present grid (e.g. Burrows et al.~2001). 

\subsection{HR 8799c}
\label{sec-hr8799c}

\begin{figure}[ht]
\centering
\includegraphics[width = \textwidth]{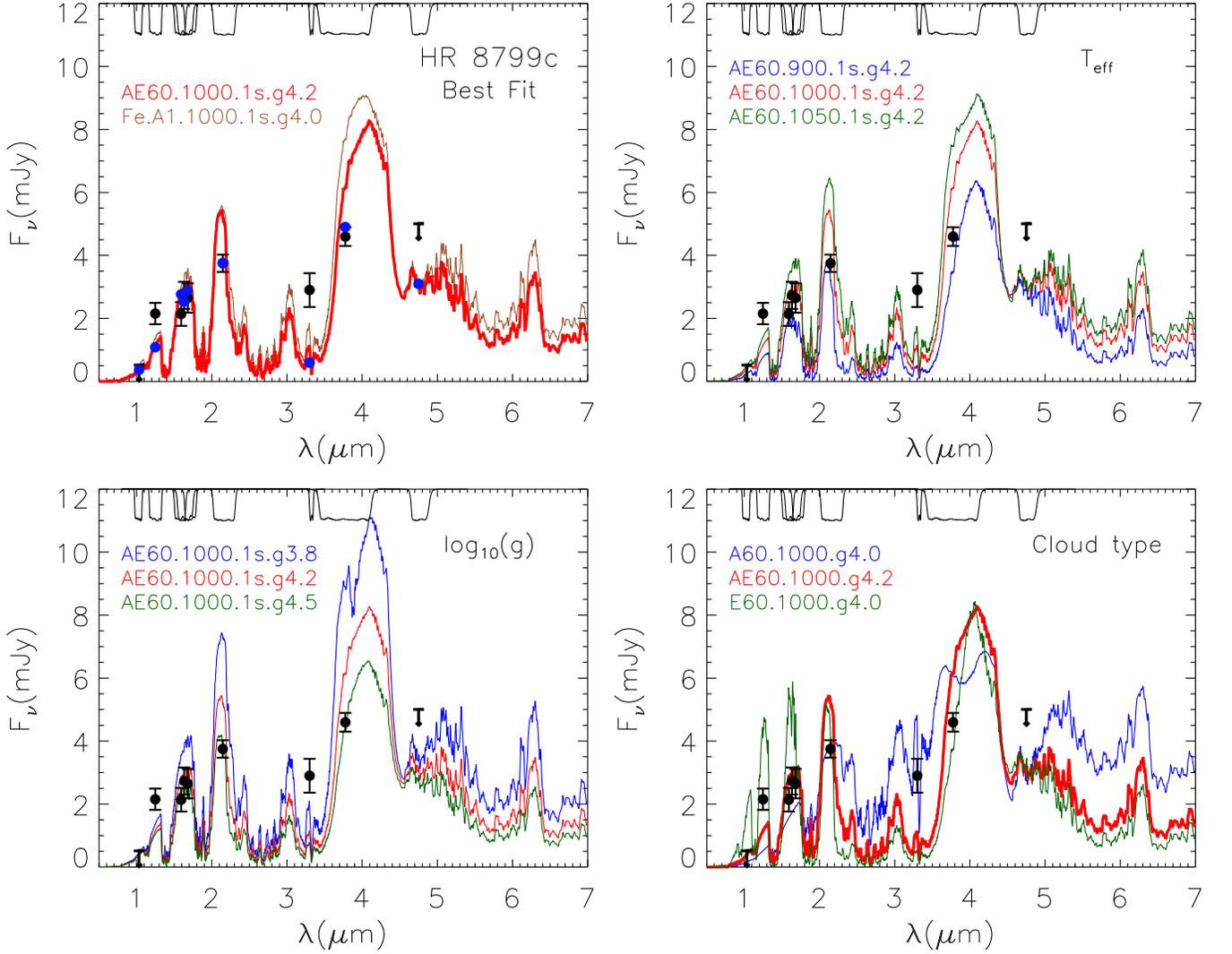}
\caption{Model spectra of HR~8799c. The observations are shown in black. The top-left panel shows the best fitting 
models. The model nomenclature is described in Fig.~\ref{fig:clouds} and in section~\ref{sec-hr8799b}. AE-type clouds with 60-$\micron$ modal particle size (see section~\ref{sec:methods}) provide a good match to the observations, similar to the case of HR~8799b shown in Fig.~\ref{fig:spectra_b}. The blue filled circles show the AE model integrated over the photometric 
bandpasses. Iron clouds, with 1\% supersaturation and 1-$\micron$ modal particle size, also explain the observations fairly well.  The best fitting AE models require $T_{\rm eff} \sim$ 1000 K and $\log_{10}(g) \sim 4.2$; current data are not sensitive to changes in metallicity. The remaining three panels show the dependence of model spectra on the key atmospheric parameters, $T_{\rm eff}$, $\log_{10}(g)$, and cloud type. T$_{\rm eff} > 1000$ K leads to substantially higher 3.8-$\micron$ flux than observed, whereas the J-band observation rules out models with lower $T_{\rm eff}$. See section~\ref{sec-hr8799c} in text for details.} 
\label{fig:spectra_c}
\end{figure}

Our model fits to observations of HR 8799c require cloud properties that are similar to 
those of HR~8799b, but with different $T_{\rm eff}$ and $\log_{10}g$. Figure~\ref{fig:spectra_c} 
shows model spectra of HR 8799c and their dependence on the model parameters. An AE-type 
forsterite cloud model with $T_{\rm eff}$ = 1000 K and $\log_{10}g$ = 4.2 provides a good match 
to the observations. The observations can also be explained with Fe clouds with 1-$\micron$ 
modal particle size and 1\% supersaturation, but with similar $T_{\rm eff}$ and $\log_{10}g$ 
of 1000 K and 4.0, respectively. The dependence of the best-fit model spectrum on 
$T_{\rm eff}$, $\log_{10}g$, and cloud cover follow the general trends described for 
HR~8799b above. 

\begin{figure}[ht]
\centering
\includegraphics[width = 0.75\textwidth]{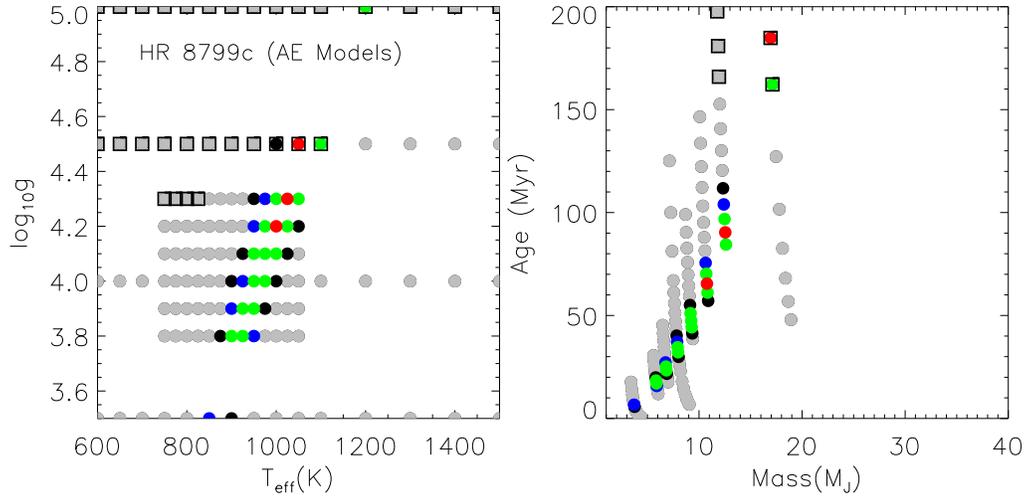}
\caption{Model space explored for HR~8799c. See Fig.~\ref{fig:cplot_all_b1} for description of the symbols 
and colors.} 
\label{fig:cplot_all_c1}
\end{figure}

\begin{figure}[ht]
\centering
\includegraphics[width = 0.75\textwidth]{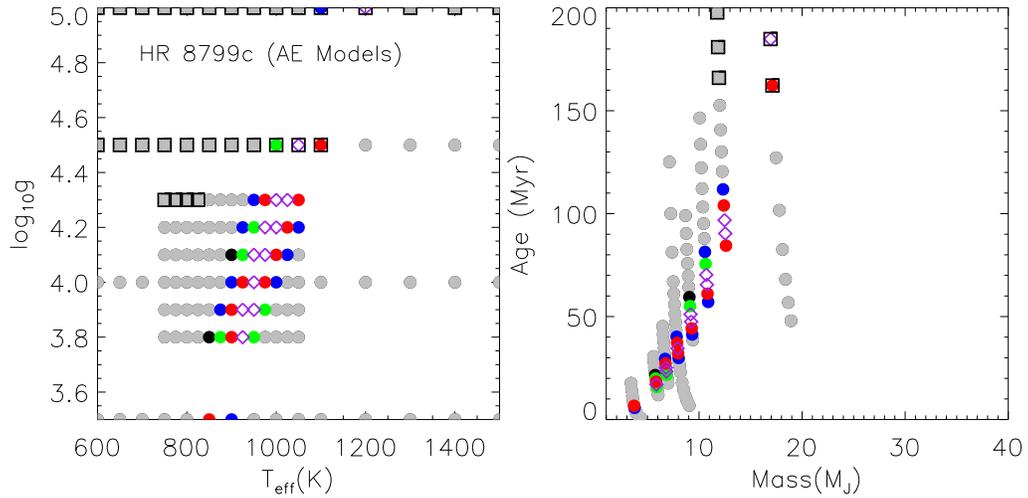}
\caption{Optimal constraints on the atmospheric parameters of HR~8799c. The symbols and colors are described in Fig.~\ref{fig:cplot_all_b1}. The parameter space explored is same as that in Fig.~\ref{fig:cplot_all_c1}. However, in the constraints shown in this figure we exclude the contribution of the 3.3-$\micron$ data point to the $\chi^2$, for reasons discussed in sections~\ref{sec-hr8799b} and \ref{sec-hr8799c}.} 
\label{fig:cplot_all_c2}
\end{figure}

Figures~\ref{fig:cplot_all_c1} and \ref{fig:cplot_all_c2} show the grid of models explored for 
HR~8799c color-coded with the level of fit to the data. We find that considering all the nine 
observations yields large values of $\chi^2$, with no model in the grid fitting to within 3$\sigma$ confidence. As with HR~8799b, we find that the observed 3.3-$\micron$ data point contributes the most to the $\chi^2$. The flux observed in the 3.3-$\micron$ band is higher than the model predictions in this bandpass 
by over 2-$\sigma$, and suggests methane depletion in HR~8799c, 
as also suggested for HR~8799b. Consequently, following the approach discussed in 
section~\ref{sec-hr8799b}, we assess the model fits without the 3.3-$\micron$ point, as shown in Fig.~\ref{fig:cplot_all_c2}, and quote the fit parameters accordingly. 

The constraints on the model parameters at different levels of confidence are shown in Fig.~\ref{fig:cplot_all_c2} 
and Table~\ref{tab:spectra_2}. At 3$\sigma$ confidence, the solutions for HR~8799c yield $T_{\rm eff}$ = 
925 - 1025 K, $\log_{10}g$ = 3.8 - 4.3, M = 6 - 13 M$_{\rm J}$, and t = 20 - 100 Myr. A representative best-fit model has $T_{\rm eff}$ = 1000 K, $\log_{10}g$ = 4.2, 
M = 11 M$_{\rm J}$, and t = 65 Myr, as shown in Table~\ref{tab:spectra_1}. In principle, an equally good 
fit is also obtained for  $T_{\rm eff}$ = 1050 K, and $\log_{10}g$ = 4.5, as shown in Fig.~\ref{fig:cplot_all_b2} 
in the purple diamond bounded by a black square, yielding a mass of 17 M$_{\rm J}$. But we rule out the 
latter solution since the corresponding age of the object would be 180 Myr, which is larger than the maximum 
age of 160 Myr estimated for the host star. 

\subsection{HR 8799d}
\label{sec-hr8799d}
\begin{figure}[ht]
\centering
\includegraphics[width = \textwidth]{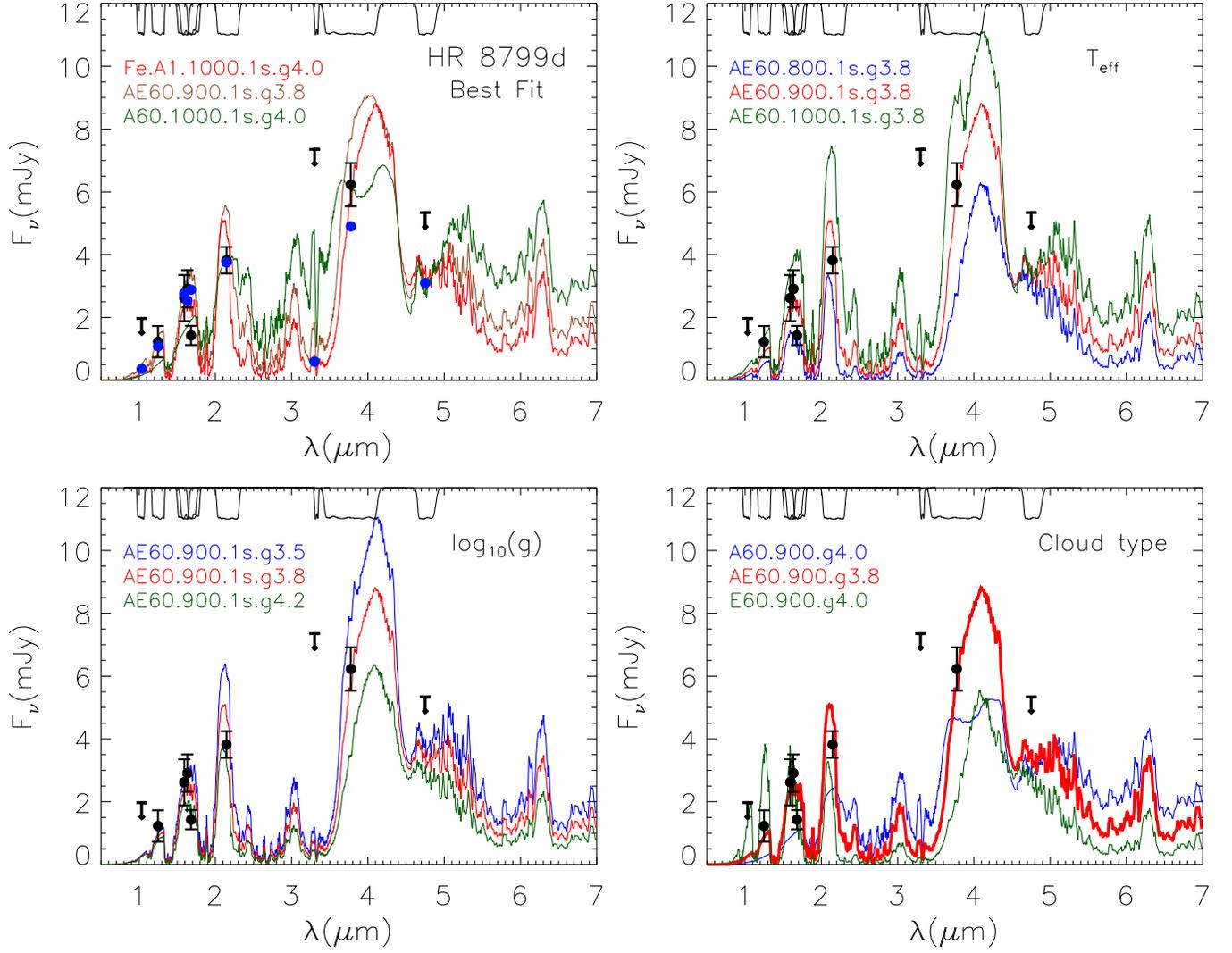}
\caption{Models fits to observations of HR~8799d. The observations are shown in black. Three 
best fitting models with different cloud types are shown in the top-left panel. The model nomenclature is described in Fig.~\ref{fig:clouds} and in section~\ref{sec-hr8799b}. The observations can be explained by a wider range of models than those for HR~8799 b and c. Nevertheless, thick clouds are required to explain the observations. As shown in the top-left panel, models with A-type clouds as well as AE-type clouds provide almost equally good fits to the data. The red curve in the top-left panel shows a best-fit model with AE-type cover, with T$_{\rm eff} \sim 900$ K and $\log_{10}(g) \sim 3.8$; the blue filled circles show the flux integrated in the photometric bandpasses. Iron clouds, with 1-$\micron$ modal particle size and 1\% supersaturation, also provide a good match to the observations, as observed for HR~8799 b and c. The remaining panels show the dependence of the spectra on different atmospheric parameters.} 
\label{fig:spectra_d}
\end{figure}

The observations of HR~8799d can be explained by a broad set of model parameters. 
Cloudless models are still inadequate to explain the data, as with HR~8799b and HR~8799c.
However, a wider range of cloudy models fit the observations. As shown in the top-left 
panel of Fig.~\ref{fig:spectra_d}, forsterite clouds of AE type and A type can explain the observations, 
with different $T_{\rm eff}$ and $\log_{10}g$. The best-fit model with AE-type cloud cover has a 
$T_{\rm eff}$ = 900 K and $\log_{10}g$ = 3.8 as shown in Table~\ref{tab:spectra_1}. On the other 
hand, a model with A-type cloud cover with $T_{\rm eff}$ and $\log_{10}g$ of 1000 K and 4.0, respectively, 
also provide a reasonable match to the data which comprises of three upper-limits and are, hence, less  
constraining. The degeneracy in cloud type leads to multiple estimates of the best-fit mass and age of HR~8799d. 
The best-fit model with AE-type clouds yields $M \sim 6$ M$_{\rm J}$ and $t$ $\sim$ 20 Myr, whereas, 
that with A-type clouds yields $M \sim 8$ M$_{\rm J}$ and $t$ $\sim$ 30 Myr. With this level of 
degeneracy, the observations are insensitive to slight changes in metallicity or to the presence 
of non-equilibrium chemistry. The observations can also be explained by models with Fe 
clouds, 1-$\micron$ modal particle size, 1\% supersaturation, and $T_{\rm eff}$ and $\log_{10}g$ 
of 1000 K and 4.0, respectively. The dependence of the best-fit models on the model parameters is 
shown in Fig.~\ref{fig:spectra_d}. 

\begin{figure}[ht]
\centering
\includegraphics[width = 0.75\textwidth]{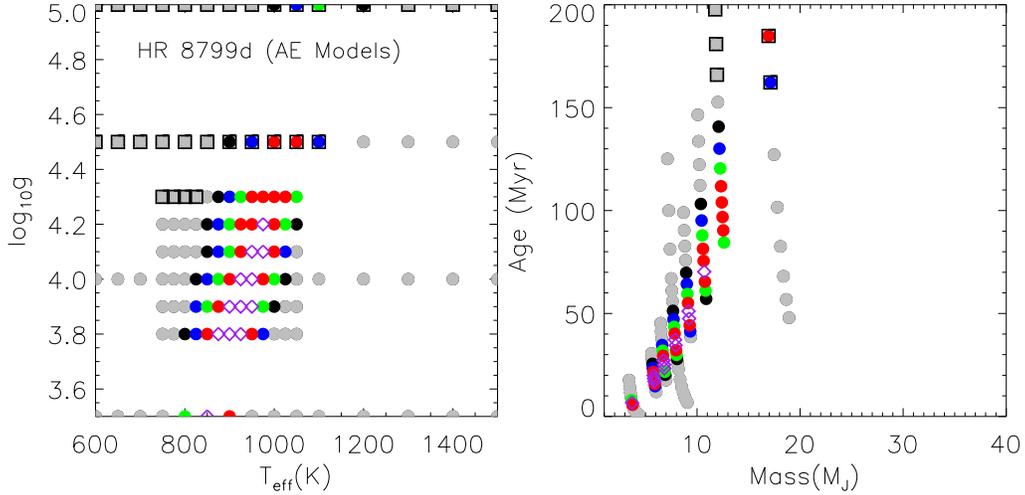}
\caption{Optimal constraints on the atmospheric parameters of HR~8799d. The approach is described in Fig.~\ref{fig:cplot_all_b1} and 
Fig.~\ref{fig:cplot_all_b2}. For HR~8799d, inclusion of the 3.3 $\micron$ photometric point in the $\chi^2$ does not significantly alter the $\chi^2$, since only an upper limit was observed in this bandpass which is higher than all the models explored.} 
\label{fig:cplot_all_d1}
\end{figure}

The range of fitting AE models for HR~8799d is shown in Fig.~\ref{fig:cplot_all_d1} at different levels 
of fit, and the 3$\sigma$ constraints are reported in Table~\ref{tab:spectra_2}. At 3$\sigma$ confidence, 
the ranges in the different physical parameters are: $T_{\rm eff}$ = 850  - 975 K, $\log_{10}g$ = 3.5 - 4.2, $M$ = 3 - 11 M$_{\rm J}$, and t = 10 - 70 Myr. We note that, unlike HR 8799b and HR 8799c, the presence 
or absence of the 3.3-$\micron$ point does not significantly influence the $\chi^2$ for model fits to the 
HR~8799d data. This is because for HR~8799d only an upper limit was observed at 3.3 $\micron$, 
which is higher than the flux predicted in this channel by all the fitting models. 

\section{New Models for Thick Cloudy Atmospheres: Predictions for 1--5 $\mu M$ spectra}
\label{sec:new_models}
Our model fits to observations of the HR~8799 planets hint at a new regime in substellar 
atmospheres. The clouds required to explain the observations of HR 8799b, c, and d, have much 
thicker physical extent compared to those traditionally required to explain the observations 
of L and T brown dwarfs at similar effective temperatures. The high-altitude clouds 
we infer in the HR~8799 planets might be a result of their lower gravities ($\log_{10}g \sim 3.5 - 4$), 
compared to most measured L and T dwarfs (Burrows et al. 2006), which can lead to slower gravitational 
settling of condensate particles (e.g. Helling 2011). Additionally, the thick clouds might also be indicators 
of high metallicity and significant vertical mixing, which might be constrained by future data. 

The observed fluxes of HR~8799b, c, and d in the near-infrared wavelengths (in z, J, H, and Ks bands) 
are two to five times lower compared with model spectra of T dwarfs at similar $T_{\rm eff}$. Major direct imaging 
programs currently underway -- e.g. IDPS and Gemini/NICI (Marois et al. 2011, in progress; 
Liu et al. 2010) -- and upcoming programs like GPI and SPHERE (MacIntosh et al. 2008; 
Beuzit et al. 2008) will likely image many new planets with a wide range of ages and masses, 
which translate into a wide range in $T_{\rm eff}$ and $g$ (e.g. Burrows et al. 1997, 2001). Because 
the mid-IR portion of a planet's SED is strongly affected by $T_{\rm eff}$ and $g$, and the presence 
of clouds, a wide spread in photometric colors and magnitudes can be expected from future detections. 

We are, therefore, motivated to provide a suite of model spectra in order to aid in the 
characterization of objects like the HR8799 planets likely to be detected in the near future. 
We tentatively refer to such objects as those bearing `thick-cloud' atmospheres, and give them the 
AE-type cloud properties required to explain the HR 8799 objects. Such models might also help 
in the characterization of objects like the cloudy T Dwarf Ross~458C  (Burgasser et al. 2010). 
We present in Table~\ref{tab:colors} models of thick-cloud type atmospheres over a 
representative grid in $T_{\rm eff}$ and $\log_{10}g$. For each 
model in the grid, we provide mass ($M$), age ($t$), and absolute magnitudes in seven photometric 
bandpasses in the infrared (z$^\prime$, Y, J, H, Ks, L$^\prime$, and M). The model grid spans 
$T_{\rm eff}$ = 600 to 1700 K, and $\log_{10}g$ = 3.5 to 5.0, 
where $g$ is in cm s$^{-2}$. The corresponding masses range between 2 and 40 M$_{\rm J}$, and ages 
range between  4 Myr and 5 Gyr. We assume the atmospheres to be in chemical equilibrium with 
solar abundances. 

\subsection{Correlations between $g$/$T_{\rm eff}$ and Mass/Age}

Low-mass objects of the thick-cloud type occupy a distinct region in atmospheric phase space. 
As shown in Table~\ref{tab:colors}, objects with masses below 10 M$_{\rm J}$ are characterized 
by $T_{\rm eff}$ $\lesssim 1100$ K, $\log_{10}g$ $\lesssim 4.5$. On the other hand, higher mass 
objects ($M \gtrsim 15$ M$_{\rm J}$) are typically associated with higher specific gravities and/or higher 
$T_{\rm eff}$. This correlation can be understood by considering the dependence of mass on 
$T_{\rm eff}$ and $g$ (Burrows et al. 2001). Mass has an approximate power-law dependence on both 
$g$ and $T_{\rm eff}$, the dependence on $g$ being steeper. For example, from Burrows et al. (2001),  

\begin{equation}
M \sim 35~{\rm M}_J~\Big(\frac{g}{10^5 {\rm cm/s^2}}\Big)^{0.64}\Big(\frac{T_{\rm eff}}{1000~{\rm K}}\Big)^{0.23}.
\label{eqn:mass}
\end{equation}
The age, on the other hand, has a power-law dependence on $g$, and a steeper inverse power-law 
dependence on $T_{\rm eff}$. 
\begin{equation}
t \sim 1~{\rm Gyr}~\Big(\frac{g}{10^5 {\rm cm/s^2}}\Big)^{1.7}\Big(\frac{1000~{\rm K}}{T_{\rm eff}}\Big)^{2.8}.
\label{eqn:age}
\end{equation}
Consequently, young planetary objects are likely to have high $T_{\rm eff}$ and low surface 
gravities.  These scaling relations justify our choice of focusing on the subset of our model space 
with log(g) $\le$ 5. 

The effect of $T_{\rm eff}$ and $g$ on the emergent spectra are apparent from the infrared 
magnitudes. As shown in Table~\ref{tab:colors} and in Fig.~\ref{fig:spectra_b}, higher $T_{\rm eff}$ 
and lower $g$ result in systematically higher fluxes in all the bandpasses. However, as shown in 
Fig.~\ref{fig:spectra_b}, the influence is highest in the Ks and L$^\prime$ bands, which are not saturated 
by water vapor absorption. 

\subsection{Comparisons with Colors and Magnitudes of L/T-Type Substellar Objects} 
To compare the predicted near-IR properties of objects with thick clouds with those 
with more standard cloud prescriptions, we generally follow the presentation 
structure in Currie et al. (2011).  We include color-magnitude loci for Case E standard 
cloud deck models from Burrows et al. (2006) for $\log_{10}g$ = 4 and 4.5 and a solar 
abundance of metals. To bracket the range in surface gravities for planet-mass objects 
from our AE model grid, we overplot color-magnitude loci for the $\log_{10}g$ = 3.5 and 4.5 cases. 

For empirical comparisons, we primarily use the L/T dwarf sample compiled by Leggett et al. (2010).  
To this sample, we add photometry for HR 8799 presented in Currie et al. (2011) 
and photometry for other planet-mass objects and very low-mass brown dwarfs 
with T$_{eff}$ = 800--1800K: 2M 1207b ($\sim$5 M$_{J}$), 1RXJ1609.1-210524 ($\sim$9 M$_{J}$), 
AB Pic ($\sim$13.5 M$_{J}$), and HD 203030b ($\sim$25 M$_{J}$) (Chauvin et al. 2004, 2005; Mohanty et al. 2007; 
Lafreniere et al. 2008, 2010; Ireland et al. 2010; Metchev and Hillenbrand 2006).

\begin{figure}[ht]
\centering
\plottwo{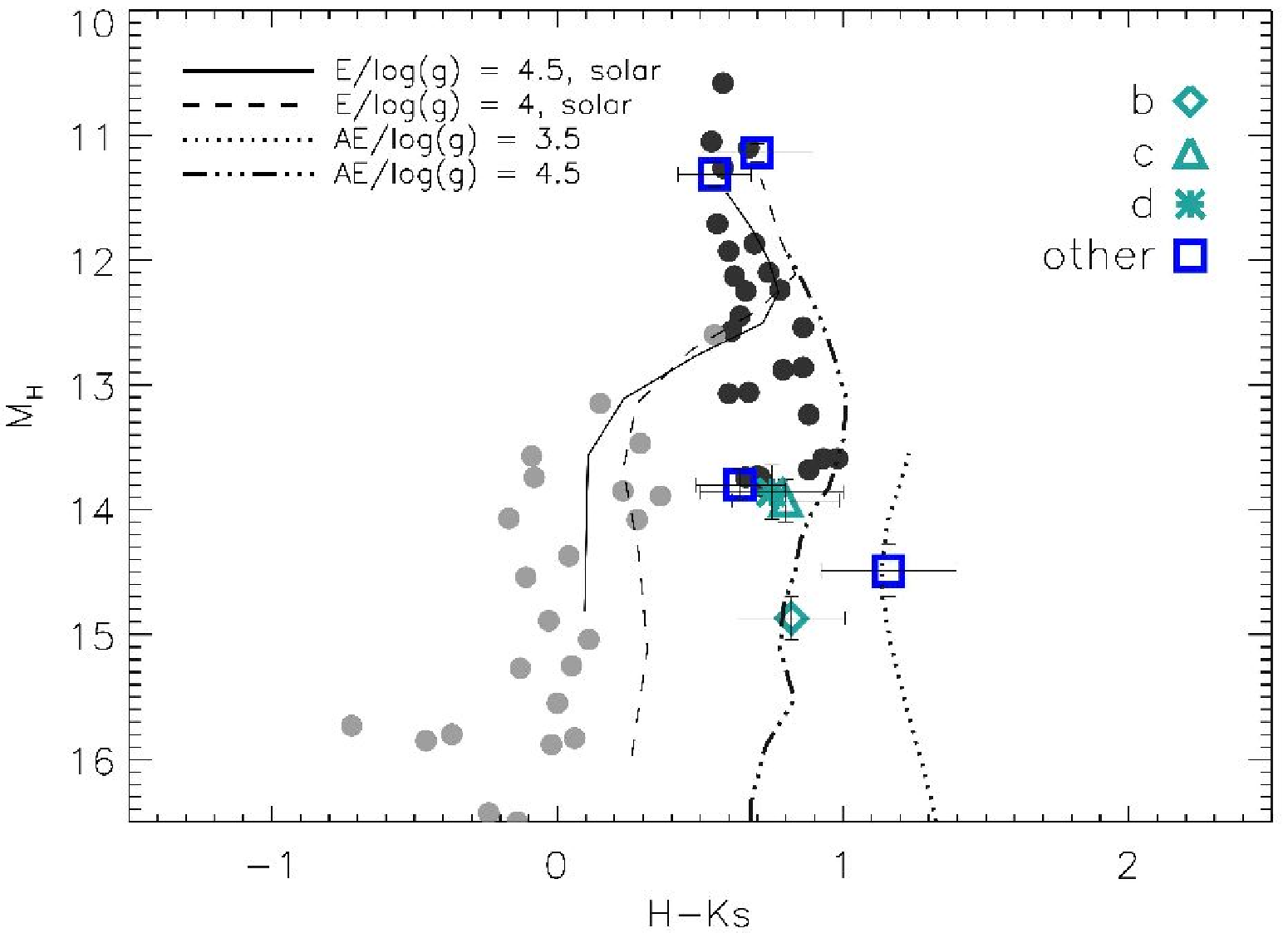}{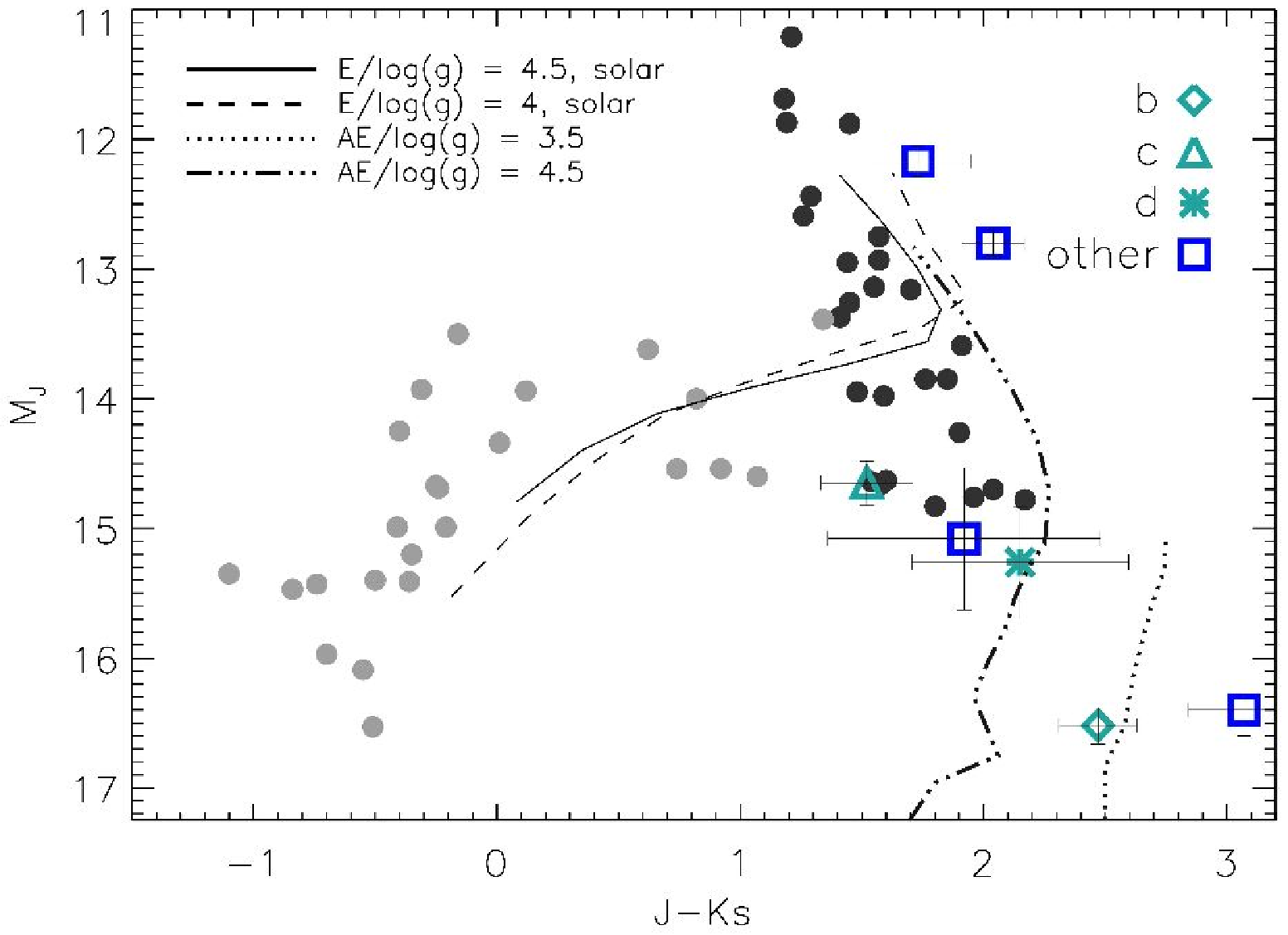}
\plottwo{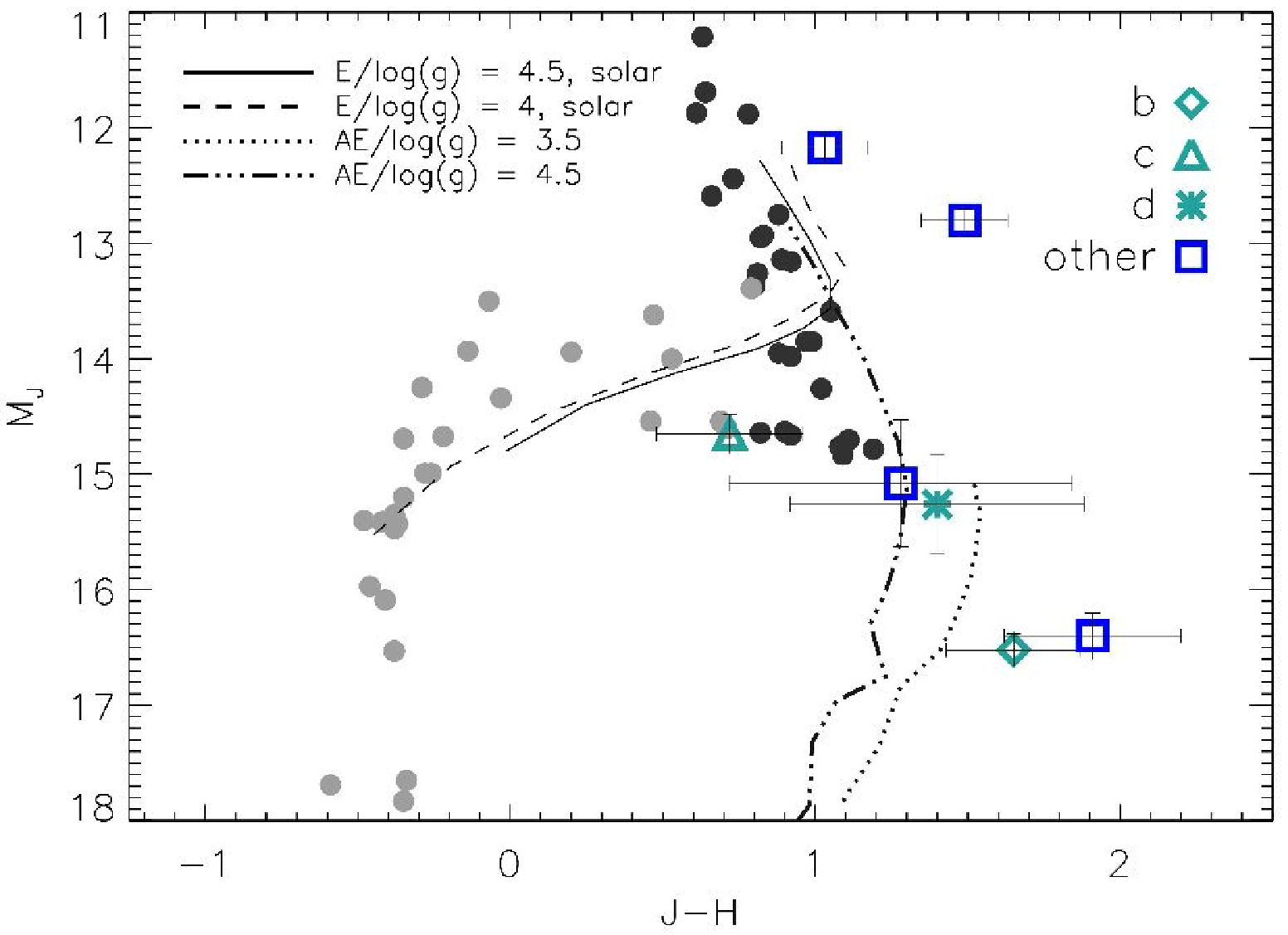}{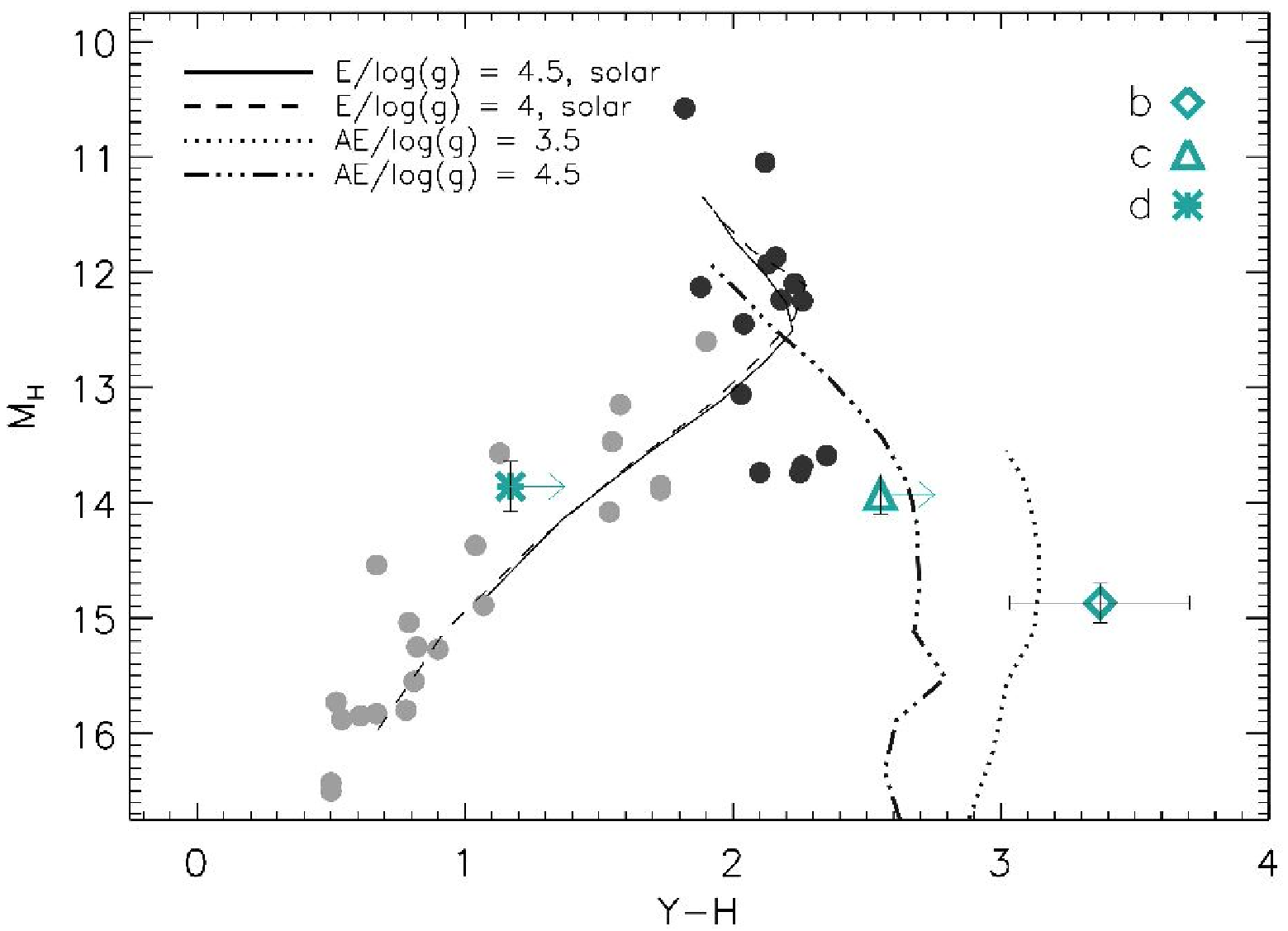}
\caption{Color-magnitude diagrams comparing the loci for standard brown dwarf atmosphere models with cloud decks (Case ``E") 
and loci for planet-mass objects with thick clouds (Case ``AE").  Overplotted are data for HR 8799 (Currie et al. 2011), 
other planet-mass objects, and field L/T dwarfs from Leggett et al. (2010), shown in grey filled circles. In all diagrams, the thick cloud models predict a far different distribution of colors and magnitudes for cool, substellar objects.  We predict that many 1--10 M$_{J}$ planets that will be discovered with GPI and SPHERE will occupy regions of color-magnitude space 
coincident with the thick cloud model locus and define a ``planet locus" distinguishable from the L/T dwarf sequence.}
\label{fig:cmd}
\end{figure}

Figure~\ref{fig:cmd} presents our color-magnitude diagrams -- H/H-K$_{s}$, J/J-K$_{s}$, J/J-H, and H/Y-H.  
In all four panels, the locus for the thick cloud models follows an extension of the L dwarf 
sequence to systematically fainter absolute magnitudes and redder colors than the standard 
cloud deck models.  Both the cloud deck and thick cloud layer models loci pass through/near the 
color-magnitude space for L dwarfs and the hottest, youngest planet-mass objects 
(1RXJ1609.1-210524 and AB Pic), though the latter's slightly redder colors are somewhat better 
matched by the thick cloud models.  However, the standard models provide far 
better matches to the colors and magnitudes of T dwarfs.  Conversely, the thick cloud layer models 
reproduce the photometry of some objects near the L/T dwarf transition, but provide far better 
matches to the photometry for faint planet-mass objects/low-mass brown dwarfs, such as the 
HR 8799 planets, HD 203030b, and 2M 1207b.

The differences between the models are most striking at temperatures less 
than $\sim$1300 K, where standard cloud deck models substantially differ by 
predicting that progressively fainter/cooler substellar objects have bluer near-IR colors.  For instance, 
for an object with  $\log_{10}g$ = 4.5, $T_{\rm eff}$ 
= 1200 K ($\approx$ 17 M$_{J}$) the predicted position in J/J-K$_{s}$ is $\approx$ [14, 1] 
in the standard cloud deck case, but [14.3, 2.2] for the Model AE case.  At lower temperatures 
($T_{\rm eff}$ $\lesssim$ 1000 K) corresponding to objects with masses approaching or below the deuterium-burning 
limit, the differences are even more pronounced.  

\subsection{Prediction of a Separate ``Planet Locus" Identifiable from Upcoming Imaging Surveys}

Our models predict that objects with thick clouds and temperatures less than $\approx$ 1300 K 
will occupy far different regions of near-IR color-magnitude diagrams than those with clouds more 
characteristic of field brown dwarfs. As inferred from Table~\ref{tab:colors}, objects with $T_{\rm eff}$ $\lesssim$ 
1300 K and  $\log_{10}g$ = 3.5 and 4.5 will have masses less than $\sim$4 M$_{J}$ and $\sim$18 M$_{J}$, 
respectively. The ages corresponding to objects with these ranges in surface gravity and temperature are 
$\gtrsim$1 Myr and $\gtrsim$117 Myr, respectively.

Upcoming imaging surveys like GPI are sensitive to detecting 1--10 M$_{J}$ around 
numerous stars with ages $\sim$ 10--100 Myr and, thus, will probe a range of 
temperatures and surface gravities where cloud deck and thick cloud models
predict very different near-IR SEDs. Therefore, our analysis predicts that the near-IR 
SEDs of many soon-to-be imaged planets will not track the L/T dwarf sequence. Rather, 
like the HR 8799 planets and 2M 1207b, they will track a distinguishable ``planet locus" 
following an extension of the L dwarf sequence, diverging from the L/T dwarf sequence 
for field brown dwarfs at the L/T transition, and extending to systematically fainter and redder 
color-magnitude positions.

\section{Conclusions}
\label{sec:conclusions}

We have generated new models to help characterize the atmospheres 
of the directly-imaged planets HR~8799b, c, and d, based on observations of broadband 
infrared photometry. Recent observational efforts have led to good data 
on the HR~8799 system. However, explaining the observations with existing atmospheric 
models had thus far proven challenging. In the discovery paper, Marois et al. (2008) 
found that spectral fits to the photometric observations yielded very high $T_{\rm eff}$, or 
unrealistically small radii, for the objects, inconsistent with the values obtained from fits to 
cooling tracks. Subsequent follow-up observations (Bowler et al. 2010, Hinz et al. 2010, 
Janson et al. 2010, Currie et al. 2011) have all reported the inadequacy of existing models 
in simultaneously reconciling all the available observations. 

The new models of thick-cloud atmospheres we have generated in this work provide good fits 
to the observations of HR~8799b, c, and d. Several recent studies on the atmospheres of these 
objects had all suggested the requirement of thick clouds to explain the observations (Marois et al. 
2008, Bowler et al. 2010, Currie et al. 2011). However, the lack of detailed models spanning different 
cloud parameters (physical thicknesses, cloud compositions, and grain size) had been an impediment 
to determining the nature of clouds required to fit the data. In a companion paper to this study, 
Currie et al. (2011) combined their new data with data from Marois et al. (2008) and a re-reduction 
of MMT data from Hinz et al. (2010) to identify models with intermediate cloud cover that might provide better fits to the observations. Their conclusions were based on our preliminary `patchy' cloud models, derived using
the modeling procedures found in Burrows et al. (2006). The present paper is meant to be a more detailed 
follow-on study to constrain the atmospheric parameters of the HR~8799 planets 
and to provide new models for future use. 

Our primary finding is that a thicker cloud cover than that required to explain observations of L 
and T brown dwarfs (e.g., the `E' model of Burrows et al. 2006) is essential to explain the present 
observations of HR~8799b, c and d. Our best-fit solutions follow from an exploration of cloudy models with different 
physical extents, cloud compositions, and modal particle sizes. In terms of the physical extent, 
we consider four cloud types: E, A, AE, and AEE. While E-type 
clouds truncate at altitude, A-type clouds extend all the way to the top of the atmosphere. 
The best-fit models for HR~8799b and c require clouds intermediate between the A and E models, 
denoted as AE models in the present work; HR~8799d allows solutions with both AE and A models. 
Models with AE type forsterite clouds with a modal particle size of 60 $\micron$ were able to explain 
the observations of all the objects. However, A-type clouds made of pure Fe with 1-$\micron$ particle 
size and 1\% super saturation could also explain the observations reasonably well, implying multiple 
solutions for the cloud composition and particle sizes that can yield the required opacities. Despite the similar 
fits obtained with the different cloud compositions, we note that the opacity spectra of the forsterite and Fe clouds 
are significantly different. Future observations in the visible and mid-IR might be able to distinguish between 
the two compositions. While forsterite clouds would be expected to show silicate features 
near $\sim$10 $\micron$ and $\sim$20 $\micron$, Fe clouds have notable opacity in the visible. 

We derive constraints on $T_{\rm eff}$, $g$, mass ($M$), and age ($t$) for HR~8799b, c, and d, based on 
spectral fits to previously published broadband photometric data in the near-infrared. At 3-$\sigma$ confidence, the constraints for HR~8799b are $T_{\rm eff}$ = 750  - 850 K, $\log_{10}g$ = 3.5 - 4.3, $M$ = 2 - 12 M$_{\rm J}$, and t = 10 - 150 Myr. The parameters of the best-fit solution are 
$T_{\rm eff}$ = 850 K, $\log_{10}g$ = 4.3, $M$ $\sim$ 12 M$_{\rm J}$, and t $\sim$ 150 Myr. For HR~8799c, the 3-$\sigma$ ranges are $T_{\rm eff}$ = 925 - 1025 K, $\log_{10}g$ = 3.8 - 4.3, $M$ = 6 - 13 M$_{\rm J}$, and t = 20 - 100 Myr. The best-fit solution has $T_{\rm eff}$ = 1000 K, $\log_{10}g$ = 4.2, $M$ $\sim$ 11 M$_{\rm J}$, and t $\sim$ 65 Myr. Finally, for HR 8799d, the 3-$\sigma$ constraints are $T_{\rm eff}$ = 850  - 975 K, $\log_{10}g$ = 3.5 - 4.2, $M$ = 3 - 11 M$_{\rm J}$, and t = 10 - 70 Myr, with the best fit obtained for $T_{\rm eff}$ = 900 K, $\log_{10}g$ = 3.8, $M$ $\sim$ 6 M$_{\rm J}$, and t $\sim$ 20 Myr. The best-fit ages for the three objects lie between 20 and 150 Myr. The age of the star has been reported to be 30 -- 160 Myr (Marois et al. 2008), which is consistent with the age ranges of the objects, suggesting that the objects might have formed in situ, and that they are coeval.  The new cloudy models developed in this work yield slightly improved fits over those previously reported by Currie et al. (2011) for the same data, and our constraints on $T_{\rm eff}$ and $\log_{10}g$, though slightly lower, are generally similar. 

We note that the observations at  3.3 $\micron$ for HR~8799b and HR~8799c indicate fluxes that are higher, by over 2-$\sigma$, 
than those predicted by the best-fitting models. This difference can in principle be explained by a lower methane abundance 
than the equilibrium abundances considered in our cloudy models. 

Motivated by our results on the HR~8799 companions, we have provided a new suite of cloudy 
atmospheric models in anticipation of the detection of similar objects in the future. These models 
are characterized by AE-type clouds that were able to fit the observations of HR~8799b, c, and d. We 
reported model spectra, photometric magnitudes in the infrared, and physical properties of the 
corresponding thick-cloud models. This model grid spans $T_{\rm eff}$ = 600 - 1700 K, $\log_{10}g$ = 3.5 - 5.0, $M$ = 2 - 40 M$_{\rm J}$, and $t$ = 4 Myr - 1 Gyr. The models assume solar abundances and 
chemical equilibrium for the gas-phase species. The models predict that objects like the HR 8799 
planets will track a distinctly different locus in color-magnitude  space 
from those of the L/T dwarf sequence. The objects would diverge in color-magnitude space from the 
L/T dwarf sequence at the L/T transition, and extend to systematically fainter and redder regions. We 
hope that the present models will help interpret the new observations of directly-imaged substellar 
objects anticipated in the near future, to characterize their masses and ages, and constrain their 
planetary nature (Spiegel et al. 2011). The grid of models reported in this work, and those used in 
Currie et al. (2011), can be obtained in electronic format at the following URL: \url{http://www.astro.princeton.edu/~burrows/}.

\acknowledgements{The authors acknowledge support in part under NASA ATP grant
NNX07AG80G, HST grants HST-GO-12181.04-A and HST-GO-12314.03-A, and
JPL/Spitzer Agreements 1417122, 1348668, 1371432, and 1377197. We thank 
David Spiegel, Mike McElwain, Ivan Hubeny, and Brendan Bowler for helpful discussions.}

\begin{deluxetable*}{c c c c c c c c c}
\tablewidth{\textwidth}  
\tabletypesize{\scriptsize}
\tablecaption{Physical parameters of best-fit models of the HR 8799\tablenotemark{a} planets}
\tablehead{\colhead{Object} & \colhead{$T_{\rm eff}$ (K)} & \colhead{$\log_{10}(g)$\tablenotemark{b}} 
& \colhead{Age (Myr)} & \colhead{Mass (M$_{\rm J}$)} & \colhead{Seperation (AU)\tablenotemark{c}} 
& \colhead{Model Name\tablenotemark{d}}}
\startdata  \\
HR 8799b & $\sim$ 850 & $\sim$ 4.3 & $\sim$ 150 & $\sim$ 12 & 68 & AE60.850.1s.g4.3\\\\
HR 8799c & $\sim$ 1000 & $\sim$ 4.2 & $\sim$ 65 & $\sim$ 11 & 38  & AE60.1000.1s.g4.2\\\\
HR 8799d & $\sim$ 900 & $\sim$ 3.8 & $\sim$ 20 & $\sim$ 6 & 24  & AE60.900.1s.g3.8\\\\
\enddata
\tablenotetext{a}{ Stellar parameters: [Fe/H] = -0.47, Age = 30 - 160 Myr, Spectral type = A5V, 
Mass = 1.5 $\pm$ 0.3 M$_\odot$ (Marois et al. 2008)}
\tablenotetext{b}{ $g$ is in cm/s$^2$. }
\tablenotetext{c}{ Projected orbital separation (Marois et al. 2008)}
\tablenotetext{d}{The nomenclature `AE60.850.1s.g4.3' denotes AE-type cloud model, 
with 60-$\micron$ modal particle size, $T_{\rm eff} = 850$ K,  1$\times$solar metallicity, 
and $\log_{10}g = 4.3$.}
\tablenotetext{e}{In cases where multiple solutions are available at similar levels of fit, a representative 
best-fit is chosen.}
\label{tab:spectra_1}
\end{deluxetable*}

\begin{deluxetable*}{c c c c c c c c}
\tablewidth{\textwidth}  
\tabletypesize{\scriptsize}
\tablecaption{Ranges of physical parameters constrained by the observations\tablenotemark{a}}
\tablehead{\colhead{Object} & \colhead{$T_{\rm eff}$ (K)} & \colhead{$\log_{10}(g)$\tablenotemark{b}} 
& \colhead{Age (Myr)} & \colhead{Mass (M$_{\rm J}$)} & \colhead{$\chi^2~$\tablenotemark{c}}}
\startdata  \\
HR 8799b & 750-850 & 3.5-4.3 & 10-150 & 2-12 & 7.2-21.1 \\ \\
HR 8799c & 925-1025 & 3.8-4.3 & 20-100 & 6-13  & 14.2-21.0\\ \\
HR 8799d & 850-975 & 3.5-4.2 & 10-70 & 3-11 & 14.9-23.5\\
\enddata
\tablenotetext{a}{The ranges for all the quantities reported here correspond to fits at 3-$\sigma$ confidence, 
as described in section~\ref{sec:observations}, and as shown in Fig.~\ref{fig:cplot_all_b2}, 
Fig.~\ref{fig:cplot_all_c2}, and Fig.~\ref{fig:cplot_all_d1}.}
\tablenotetext{a}{ $g$ is in cm s$^{-2}$. }
\tablenotetext{c}{$\chi^2$ is defined in section \ref{sec:observations}.}
\label{tab:spectra_2}
\end{deluxetable*}

\begin{deluxetable*}{c c c c c c c c c c c}
\tablewidth{\textwidth}  
\tabletypesize{\scriptsize}
\tablecaption{Physical and photometric characteristics of objects with thick ``AE type" clouds\tablenotemark{a,b}}
\tablehead{\colhead{$T_{\rm eff}$ (K)} & \colhead{$\log_{10}g$ (cm$^2$/s)} & \colhead{Mass (M$_{\rm J}$)} & \colhead{Age (Gyr)} & \colhead{z$^\prime$} & \colhead{Y} & \colhead{J} & \colhead{H} & \colhead{Ks} & \colhead{L$^\prime$} & \colhead{M} }
\startdata  
   600 &    3.5 &     2.17 &    0.023 &    21.74 &    19.62 &    17.83 &    16.74 &    15.39 &    13.73 &    12.25\\
   650 &    3.5 &     2.23 &    0.017 &    21.14 &    19.11 &    17.35 &    16.14 &    14.85 &    13.26 &    12.02\\
   700 &    3.5 &     2.30 &    0.014 &    20.57 &    18.58 &    16.84 &    15.56 &    14.34 &    12.86 &    11.85\\
   750 &    3.5 &     2.38 &    0.011 &    20.15 &    18.23 &    16.52 &    15.11 &    13.94 &    12.46 &    11.71\\
   800 &    3.5 &     2.45 &    0.009 &    19.70 &    17.85 &    16.18 &    14.71 &    13.57 &    12.11 &    11.62\\
   850 &    3.5 &     2.51 &    0.007 &    19.30 &    17.51 &    15.88 &    14.37 &    13.23 &    11.80 &    11.57\\
   900 &    3.5 &     2.58 &    0.006 &    18.93 &    17.18 &    15.60 &    14.07 &    12.91 &    11.54 &    11.56\\
   950 &    3.5 &     2.65 &    0.005 &    18.58 &    16.89 &    15.34 &    13.80 &    12.60 &    11.31 &    11.56\\
  1000 &    3.5 &     2.72 &    0.004 &    18.22 &    16.57 &    15.07 &    13.55 &    12.32 &    11.10 &    11.57\\
   600 &    4.0 &     5.85 &    0.125 &    22.02 &    19.87 &    18.11 &    17.11 &    16.02 &    14.10 &    12.60\\
   650 &    4.0 &     5.98 &    0.098 &    21.18 &    19.14 &    17.48 &    16.46 &    15.45 &    13.71 &    12.39\\
   700 &    4.0 &     6.11 &    0.078 &    20.85 &    18.87 &    17.18 &    15.98 &    14.92 &    13.30 &    12.18\\
   750 &    4.0 &     6.24 &    0.064 &    19.32 &    17.54 &    16.07 &    15.00 &    14.16 &    12.86 &    12.01\\
   800 &    4.0 &     6.38 &    0.052 &    19.89 &    18.04 &    16.42 &    15.10 &    14.09 &    12.63 &    11.89\\
   850 &    4.0 &     6.52 &    0.043 &    19.45 &    17.66 &    16.09 &    14.73 &    13.74 &    12.33 &    11.81\\
   900 &    4.0 &     6.67 &    0.035 &    19.06 &    17.32 &    15.79 &    14.41 &    13.42 &    12.06 &    11.75\\
   950 &    4.0 &     6.83 &    0.029 &    18.75 &    17.06 &    15.56 &    14.15 &    13.12 &    11.81 &    11.73\\
  1000 &    4.0 &     6.98 &    0.025 &    18.43 &    16.78 &    15.32 &    13.90 &    12.84 &    11.60 &    11.72\\
  1050 &    4.0 &     7.12 &    0.021 &    18.11 &    16.51 &    15.08 &    13.67 &    12.58 &    11.41 &    11.73\\
  1100 &    4.0 &     7.25 &    0.019 &    17.80 &    16.23 &    14.85 &    13.46 &    12.35 &    11.23 &    11.75\\
   600 &    4.5 &    14.46 &    1.005 &    22.19 &    19.98 &    18.28 &    17.42 &    16.86 &    14.46 &    12.96\\
   650 &    4.5 &    14.69 &    0.787 &    21.63 &    19.53 &    17.87 &    16.89 &    16.22 &    14.08 &    12.74\\
   700 &    4.5 &    14.93 &    0.619 &    20.91 &    18.90 &    17.32 &    16.33 &    15.65 &    13.73 &    12.54\\
   750 &    4.5 &    15.15 &    0.510 &    20.41 &    18.50 &    16.96 &    15.89 &    15.16 &    13.42 &    12.37\\
   800 &    4.5 &    15.38 &    0.434 &    20.17 &    18.31 &    16.75 &    15.52 &    14.69 &    13.09 &    12.20\\
   850 &    4.5 &    15.60 &    0.373 &    19.60 &    17.80 &    16.30 &    15.12 &    14.34 &    12.84 &    12.10\\
   900 &    4.5 &    15.83 &    0.324 &    19.23 &    17.49 &    16.02 &    14.79 &    14.00 &    12.58 &    12.01\\
   950 &    4.5 &    16.07 &    0.281 &    18.90 &    17.20 &    15.77 &    14.51 &    13.68 &    12.34 &    11.94\\
  1000 &    4.5 &    16.33 &    0.245 &    18.59 &    16.94 &    15.53 &    14.25 &    13.40 &    12.12 &    11.90\\
  1050 &    4.5 &    16.57 &    0.216 &    18.30 &    16.69 &    15.31 &    14.02 &    13.13 &    11.91 &    11.89\\
  1100 &    4.5 &    16.82 &    0.192 &    18.04 &    16.47 &    15.12 &    13.82 &    12.87 &    11.73 &    11.89\\
  1200 &    4.5 &    17.37 &    0.149 &    17.50 &    16.00 &    14.71 &    13.44 &    12.44 &    11.41 &    11.92\\
  1300 &    4.5 &    17.93 &    0.117 &    16.97 &    15.54 &    14.32 &    13.11 &    12.10 &    11.10 &    11.91\\
  1400 &    4.5 &    18.42 &    0.096 &    16.45 &    15.09 &    13.94 &    12.79 &    11.82 &    10.85 &    11.80\\
  1500 &    4.5 &    18.95 &    0.079 &    15.89 &    14.61 &    13.53 &    12.47 &    11.55 &    10.63 &    11.65\\
  1600 &    4.5 &    19.51 &    0.066 &    15.40 &    14.21 &    13.17 &    12.18 &    11.32 &    10.45 &    11.46\\
  1700 &    4.5 &    20.04 &    0.057 &    14.94 &    13.83 &    12.83 &    11.92 &    11.12 &    10.30 &    11.23\\
   600 &    5.0 &    32.78 &    4.569 &    22.46 &    20.11 &    18.45 &    17.77 &    17.88 &    14.83 &    13.36\\
   650 &    5.0 &    33.12 &    3.627 &    21.74 &    19.50 &    17.92 &    17.18 &    17.20 &    14.47 &    13.15\\
   700 &    5.0 &    33.45 &    2.938 &    21.34 &    19.23 &    17.66 &    16.73 &    16.48 &    14.09 &    12.91\\
   750 &    5.0 &    33.80 &    2.417 &    21.12 &    19.04 &    17.50 &    16.55 &    16.24 &    13.94 &    12.81\\
   800 &    5.0 &    34.16 &    2.021 &    20.32 &    18.37 &    16.90 &    15.89 &    15.48 &    13.54 &    12.57\\
   850 &    5.0 &    34.51 &    1.711 &    19.87 &    17.99 &    16.56 &    15.53 &    15.07 &    13.30 &    12.43\\
   900 &    5.0 &    34.88 &    1.463 &    19.48 &    17.68 &    16.28 &    15.21 &    14.68 &    13.06 &    12.31\\
   950 &    5.0 &    35.28 &    1.256 &    19.10 &    17.35 &    15.99 &    14.90 &    14.35 &    12.85 &    12.22\\
  1000 &    5.0 &    35.70 &    1.078 &    18.80 &    17.12 &    15.77 &    14.64 &    14.03 &    12.63 &    12.14\\
  1050 &    5.0 &    36.18 &    0.920 &    18.51 &    16.87 &    15.55 &    14.40 &    13.74 &    12.43 &    12.09\\
  1100 &    5.0 &    36.73 &    0.768 &    18.23 &    16.64 &    15.34 &    14.18 &    13.48 &    12.25 &    12.06\\
  1200 &    5.0 &    37.54 &    0.583 &    17.73 &    16.21 &    14.97 &    13.79 &    13.01 &    11.91 &    12.05\\
  1300 &    5.0 &    38.04 &    0.503 &    17.27 &    15.83 &    14.63 &    13.47 &    12.61 &    11.62 &    12.09\\
  1400 &    5.0 &    38.62 &    0.442 &    16.79 &    15.42 &    14.28 &    13.17 &    12.29 &    11.34 &    12.07\\
  1500 &    5.0 &    39.70 &    0.360 &    16.31 &    15.03 &    13.93 &    12.89 &    12.03 &    11.10 &    11.97\\
  1600 &    5.0 &    40.84 &    0.294 &    15.83 &    14.62 &    13.58 &    12.61 &    11.78 &    10.90 &    11.84\\
  1700 &    5.0 &    41.62 &    0.260 &    15.40 &    14.27 &    13.25 &    12.35 &    11.57 &    10.74 &    11.66\\
\enddata
\tablenotetext{a}{All the models are assumed to have solar metallicity.}
\tablenotetext{b}{Absolute photometric magnitudes are reported in the following filters: 
z$^\prime$ (SDSS), Y(MKO), J (MKO), H (MKO), Ks (ISAAC), L$^\prime$ (MKO), and M (MKO).}
\label{tab:colors}
\end{deluxetable*}

\end{document}